\documentclass{aa}

\usepackage{natbib}
\bibpunct{(}{)}{;}{a}{}{,}  
\usepackage{amsmath}
\usepackage{graphicx}
\usepackage{longtable}
\usepackage{multicol}
\usepackage{color}
\usepackage{lscape}
\usepackage[varg]{txfonts}
\usepackage{hyperref}

\begin{document}
   \title{Determination of robust metallicities for metal-rich red giant branch stars
   		\thanks{Based on observations made with the ESO/VLT, at Paranal Observatory, under programme 067.B-0382(A) and on data obtained from the ESO Science Archive Facility under programme 065.L-0340(A), 067.D-0489(A), and 077.B-0327(A) and from the Keck Observatory Archive under programme C53H and C19H.}}

   \subtitle{An application to the globular cluster NGC\,6528}

   \author{C. Liu
   	\inst{1,2}
          \and
          G. Ruchti\inst{1}
          \and
         S. Feltzing\inst{1}
         \and
         F. Primas\inst{3}
          }

   \institute{Lund Observatory, Department of Astronomy and Theoretical Physics,
              Box 43, SE--221 00 Lund, Sweden\\
              \email{[cheng, greg, sofia]@astro.lu.se}
              \and
              Key Lab of Optical Astronomy, National Astronomical Observatories, Chinese Academy of Sciences, A20 Datun Road, Chaoyang, Beijing 100012, China
              \and
              European Southern Observatory, D-85748 Garching, Germany\\
              \email{fprimas@eso.org}
              }

   \date{Received 05 May 2016; accepted 05 January 2017}

 
  \abstract
      {The study of the Milky Way  relies on our ability to interpret the light from stars correctly. With the advent of the astrometric ESA mission {\sl Gaia} we will enter a new era where the study of the Milky Way can be undertaken on much larger scales than currently possible. In particular we will be able to obtain full 3D space motions of red giant stars at large distances. This calls for a reinvestigation of how reliably we can determine, e.g., iron abundances in such stars and how well they reproduce those of dwarf stars.}
      {Here we explore robust ways to determine the iron content of
        metal-rich giant stars. We aim to understand what biases and shortcomings widely applied methods suffer from.}
   {In this study we are mainly concerned with standard methods to analyse stellar spectra. This includes the analysis of individual lines to determine stellar parameters, analysis of the broad wings of certain lines (e.g., H$\alpha$ and calcium lines) to determine effective temperature and surface gravity for the stars.}
   {For NGC 6528 we find that [Fe/H]  = $+0.04$\,dex with a scatter of $\sigma=0.07$ dex, which gives an error in the
derived mean abundance of 0.02 dex.}
  {Our work has two important conclusions for analysis of metal-rich red giant branch stars. 1) For spectra with S/N below about 35 per reduced pixel 
  [Fe/H] become too high, 2) Determination of $T_{\rm eff}$ using the wings of the H$\alpha$ line results in [Fe/H] values about 
  0.1\,dex higher than if excitational equilibrium is used. The last conclusion is perhaps not surprising as we expect NLTE effect to 
  become more prominent in cooler stars and we can not use the the wings of the H$\alpha$ line to determine $T_{\rm eff}$ for the 
  cool stars in our sample. We therefore recommend that in studies of metal-rich red giant stars care needs to be taken to obtain
  sufficient calibration data in order to be able to also use the cooler stars.}


   \maketitle
%

\section{Introduction}


The analysis of spectra of red giant branch stars is problematic. As
we progress along the red giant branch upwards, to cooler and lower
gravity stars a number of phenomena start appearing more and more
strongly in the spectra. In particular, the atomic lines get stronger,
causing more blends and the cooler atmospheres allow for the formation
of molecules such as TiO, CN, and MgH resulting in strong molecular
features. This makes it progressively more and more difficult to find
(or define) the continuum in the spectrum. As many methods rely on the
identification of the continuum, this causes a problem. 

Our main aim with the present investigation is to study how we
can, in a robust way, determine the iron abundance for metal-rich
giants, such as those seen in the Galactic bulge and  old,
metal-rich globular clusters. We are  interested in providing a
better understanding of the influence on the derived iron abundance
from the signal-to-noise ratio (S/N) in the analysed spectra and any
influence on the method used to determine the other stellar parameters,
e.g., $T_{\rm eff}$ and $\log g$, on the final iron abundance.

Why is this interesting? The study of the Milky Way has in the last
couple of decades become a major area in astrophysics. This is due to
two influential astrometric missions: the past Hipparcos mission
\citep{1997A&A...323L..49P} and the current {\sl Gaia} mission
\citep[][where the first paper describes the concept behind the {\sl Gaia}
  proposal to the European Space Agency and the second paper gives an
  overview of, {\sl Gaia} just before
  launch]{2001A&A...369..339P,2013arXiv1310.3485B}. The Hipparcos
catalogue enabled a new type of studies with determination of accurate
velocities and ages for samples consisting of many hundreds of stars near the sun
\citep[e.g.,][]{2004A&A...415..155B,2005A&A...438..139S,2011MNRAS.414.2893F,2012A&A...545A..32A,2013A&A...554A..44A,2014A&A...562A..71B}. These
studies were done on F and G dwarf stars. However, {\sl Gaia} will provide
data of the same quality as Hipparcos for a significantly larger
volume of the Milky Way, reaching all the way in to the Galactic
bulge. But, the stars available for spectroscopic work will (as today)
be the red giant branch stars as these are intrinsically sufficiently
bright to be possible to observe with current instrumentation at a
distance of 8 kpc.  The only option to obtain spectra for dwarf stars
in the Galactic bulge is currently to take advantage of microlensing
events \citep[e.g.,][]{2013A&A...549A.147B}. This is, however, not an
option if we want to study large samples. Thus it remains important to
study how we best can determine stellar parameters and elemental
abundances of evolved red giant stars in a robust manner. This paper
deals with this topic.

The spectroscopic and photometric data {\sl Gaia} benchmark stars and
stars in the metal-rich globular cluster NGC\,6528 are presented in
Sect.~\ref{sect:data}. Our basic tools and linelist are described in
Sect.~\ref{sect:lat}. To determine stellar parameters for red giants
we explore different methods and test them on the benchmark stars in
Sect.~\ref{sect:methodology}. In Sect.~\ref{sect:6528} we apply our
analysis to stars in NGC\,6528 and discuss problems in the analysis.
In Sect.~\ref{sect:metallicity}, we determine [Fe/H] for
NGC\,6528. Section~\ref{sect:conclusions} provides our conclusions and
discusses future developments.


\section{Description of data}\label{sect:data}

\subsection{Spectra}

\subsubsection{Benchmark stars }
\label{sect:BS}

In order to work out the best analysis techniques for
metal-rich cool giant stars we make use of spectra of stars with
well-determined stellar parameters, the so called {\sl Gaia} benchmark stars
\citep{2015arXiv150606095H}. These have a wealth of high quality data
available, in particular they have interferrometric observations
enabling a direct determination of their radii and effective
temperatures ($T_{\rm{eff}}$). As we are interested in metal-rich stars, we
selected ten red giants which have iron abundances greater than
--0.55\,dex \citep{2014A&A...564A.133J}. The ten stars, which have $T_{\rm{eff}}$ similar to the star's of NGC\,6528, are Arcturus, $\mu$Leo, HD107328, $\beta$Gem, $\epsilon$Vir, $\xi$Hya, $\alpha$Cet, $\gamma$Sge, $\alpha$Tau, and $\beta$Ara. Their reference values for the stellar parameters are listed in
Table\,~\ref{bsp}.

The fifteen spectra with high S/N ($> 250$) were taken from the ESO
archive and from observations done with NARVAL. The spectra from the
ESO archive consist of observations taken with UVES
\citep{2000SPIE.4008..534D} and HARPS \citep{2003Msngr.114...20M}. The
NARVAL spectra are further described in \citet{2014A&A...566A..98B},
which also gives details on how to retrieve the spectra from their
archive. It should be noted that five stars, $\mu$Leo, $\beta$Gem, $\xi$Hya, $\alpha$Cet, and $\gamma$Sge, have one spectrum each, while the rest of stars have two spectra observed in different instruments. For example, the star Arcturus was observed with the NARVAL and UVES spectrographs, while $\beta$Ara was observed with the HARPS and UVES spectrographs.


\subsubsection{Stars in NGC\,6528 }

\begin{table}
\setlength{\tabcolsep}{7pt}
\begin{tiny}
\caption{Basic data for NGC\,6528}
\label{mref}
\begin{tabular}{lrrr}
      \hline
      \hline
R.A. & Dec. & (m-M) & E(B-V)\\
 18 04 49.64 & --30 03 22.6 & 16.17 & 0.54 \\
           \hline
\end{tabular}
\end{tiny}
\end{table}

\begin{longtab}
\setlength{\tabcolsep}{3pt}
\begin{tiny}
\begin{longtable}{lllrrrrrrrrrr}
\caption{\label{gcs} Basic information for all stars analysed in this study. The
  first part of the table contains the data for stars observed with
  UVES on VLT in stand-alone mode, the second part lists the stars
  observed with FLAMES-UVES on VLT, and the third part stars observed
  with HIRES on KECK.} \\
\hline\hline
\#&Obj. &ID &\multicolumn{1}{l}{R.A.}&\multicolumn{1}{l}{DEC}
&$J_{0}^{a}$ &$H_{0}^{a}$ &$K_{\rm{s,0}}^{a}$
&\multicolumn{1}{l}{$v_{\rm{r}}$} &Date &Exp. &Instrument
&S/N$^{b}$\\ 
&& & & &(mag) &(mag) &(mag) &\multicolumn{1}{l}{(km s$^{-1}$)} & &(s) \\ 
\hline \endfirsthead
\caption{continued.}\\
\hline\hline
 \# &     Obj. &ID &\multicolumn{1}{l}{R.A.} &\multicolumn{1}{l}{DEC} &$J_{0}^{a}$ &$H_{0}^{a}$ &$K_{\rm{s,0}}^{a}$ &\multicolumn{1}{l}{$v_{\rm{r}}$} &Date  &Exp. &Instrument &(S/N)/px$^{b}$\\
      &        &       &         &  &(mag)    &(mag)      &(mag)     &\multicolumn{1}{l}{(km s$^{-1}$)}  &  &(s)  \\   
      \hline
      \endhead
      \hline
      \endfoot
 (01) &     01  &2MASS 18044807-3002301 &18 04 48.02  &--30 02 30  &13.11  &12.60  &12.46 &--52.46$\pm$0.18 &14.05.01 &3x4800 &UVES &39\\
 (02) &       02  &VVV J180448.98-300152.34 &18 04 48.98  &--30 01 52 &16.45  &16.09  &16.11 &12.00$\pm$0.45  &15.05.01 &3x4800 &UVES &167\\
 (03) &       03  &2MASS 18044889-3004066  &18 04 48.99  &--30 04 07 &12.37 &11.76  &11.58 &204.63$\pm$0.21 &13.06.01 &3x4800 &UVES &42\\
  (04) &      04  &2MASS 18044912-3004010  &18 04 49.27  &--30 04 01 &         &    &    &216.65$\pm$0.19    &14.05.01 &2x4800 &UVES &41\\
(05) &        05  &VVV J180450.06-300159.90  &18 04 50.20  &--30 01 59 &15.58  &15.29  &15.20 &211.68$\pm$0.22    &19.06.01 &3x4800 &UVES &55\\
(06) &        06  &2MASS 18044721-3002305  &18 04 47.17  &--30 02 31 &13.12  &12.66  &12.56 &211.09$\pm$0.29  &20.05.01 &2x5400 &UVES &29\\
(07) &         07  &VVV J180447.91-300340.06  &18 04 47.78  &--30 03 39 &13.41  &12.89  &12.79 &216.16$\pm$0.23  &13.05.01 &3x4800 &UVES &46\\
(08) &     1-02  &VVV J180451.33-300301.99 &18 04 51.23  &--30 03 03 &13.93  &13.40  &13.35 &216.09$\pm$0.34 &26.06.00 &2x4800 &UVES &28\\
(09) &     1-16  &2MASS 18044953-3003044  &18 04 49.60  &--30 03 03 &12.77  &12.21  &12.11 &208.41$\pm$0.21 &26.06.00 &2x3600 &UVES &62\\
(10) &     1-24  &2MASS 18045029-3003099  &18 04 50.37  &--30 03 10 &13.18  &12.68  &12.58 &213.62$\pm$0.28 &26.06.00 &4584 &UVES &38\\
(11) &     1-36  &   &18 04 50.92  &--30 03 48 &            &             &           &217.17$\pm$0.18 &26.06.00 &2x5400 &UVES &44\\
(12) &     1-42  &2MASS 18044900-3003264  &18 04 49.11  &--30 03 28 &12.17  &11.48  &11.33 &214.55$\pm$0.26 &26.06.00 &3x3600 &UVES &37\\
   \hline
(13) &        60  &VVV J180448.38-300255.09  &18 04 48.36  &--30 02 55 &12.17  &11.61  &11.44 &206.01$\pm$0.26  &29.06.06 &2x2700 &Flames-UVES &39\\
        &     &  &                       &                    &            &            &            &206.68$\pm$0.23  &31.05.06 &2x2700 &Flames-UVES\\
(14) &        61  &2MASS 18044812-3003082 &18 04 48.12  &--30 03 08 &9.51   &8.71   &8.44 &212.32$\pm$1.30  &29.06.06 &2x2700 &Flames-UVES &19\\
        &     &  &                       &                    &            &            &            &212.57$\pm$1.36  &01.07.06 &2x2700 &Flames-UVES\\
        &     &  &                       &                    &            &            &            &213.31$\pm$1.30  &19.07.06 &2x2700 &Flames-UVES\\
(15) &        62  &2MASS 18044578-3002470  &18 04 45.79  &--30 02 47 &11.30  &10.49  &10.34 &212.84$\pm$0.49 &29.06.06 &2x2700 &Flames-UVES &43\\
       &      &  &                       &                    &            &            &            &213.10$\pm$0.37  &01.07.06 &2x2700 &Flames-UVES\\
        &     &  &                       &                    &            &            &            &213.16$\pm$0.37  &19.07.06 &2x2700 &Flames-UVES\\      
(16) &        63  &2MASS 18044537-3003468  &18 04 45.36  &--30 03 46 &11.32  &10.63  &10.41 &209.18$\pm$0.26  &29.06.06 &2x2700 &Flames-UVES &35\\
       &      &  &                       &                    &            &            &            &209.41$\pm$0.30  &31.05.06 &2x2700 &Flames-UVES\\
(17) &        64  &2MASS 18044775-3003469  &18 04 47.75  &--30 03 47 &12.16  &11.48  &11.33 &214.69$\pm$0.60  &29.06.06 &2x2700 &Flames-UVES &38\\
        &     &  &                       &                    &            &            &            &214.77$\pm$0.47  &01.07.06 &2x2700 &Flames-UVES\\
         &    &  &                       &                    &            &            &            &214.92$\pm$0.31  &19.07.06 &2x2700 &Flames-UVES\\
(18) &        65  &2MASS 18044890-3003472  &18 04 48.90  &--30 03 47 &12.03  &11.25  &11.22 &205.08$\pm$0.26  &29.06.06 &2x2700 &Flames-UVES &57\\
       &      &  &                       &                    &            &            &            &205.29$\pm$0.23  &31.05.06 &2x2700 &Flames-UVES\\
(19) &        66  &2MASS 18044963-3003382  &18 04 49.61  &--30 03 37 &9.17   &8.43   &8.20 &215.16$\pm$1.74  &26.05.06 &3x2700 &Flames-UVES &23\\
       &      &  &                       &                    &            &            &            &216.25$\pm$1.14  &19.07.06 &3x2700 &Flames-UVES\\
        &     &  &                       &                    &            &            &            &214.70$\pm$1.36  &20.07.06 &2x2700 &Flames-UVES\\
      &      &  &                       &                    &            &            &            &216.08$\pm$1.14  &27.07.06 &2x2700 &Flames-UVES\\
(20) &        68  &2MASS 18045165-3003232  &18 04 51.64  &--30 03 23 &10.32   &9.47   &9.27  &202.52$\pm$0.57  &29.06.06 &2x2700 &Flames-UVES &18\\
       &      &  &                       &                    &            &            &            &203.38$\pm$0.54  &31.05.06 &2x2700 &Flames-UVES\\
(21) &        69  &2MASS 18045116-3003150  &18 04 51.16  &--30 03 14 &11.68  &10.98  &10.88 &218.26$\pm$0.52 &29.06.06 &2x2700 &Flames-UVES &39\\
        &     &  &                       &                    &            &            &            &218.31$\pm$0.53  &01.07.06 &2x2700 &Flames-UVES\\
        &     &  &                       &                    &            &            &            &218.17$\pm$0.44  &19.07.06 &2x2700 &Flames-UVES\\ 
(22) &        70  &2MASS 18045136-3003047  &18 04 51.36  &--30 03 04 &10.86  &10.09  &9.90 &216.06$\pm$0.63  &29.06.06 &2x2700 &Flames-UVES &26\\
        &     &  &                       &                    &            &            &            &215.82$\pm$0.45  &31.05.06 &2x2700 &Flames-UVES\\
 (23) &       71  &2MASS 18045042-3002591 &18 04 50.46  &--30 03 00 &13.53  &13.03  &12.94 &205.57$\pm$0.31  &26.05.06 &3x2700 &Flames-UVES &36\\
        &     &  &                       &                    &            &            &            &205.80$\pm$0.49  &19.07.06 &3x2700 &Flames-UVES\\
        &     &  &                       &                    &            &            &            &205.62$\pm$0.31  &20.07.06 &2x2700 &Flames-UVES\\
        &     &  &                       &                    &            &            &            &205.63$\pm$0.32  &27.07.06 &2x2700 &Flames-UVES\\
(24) &        72  &VVV J180443.90-300344.38  &18 04 43.89  &--30 03 44 &13.83  &13.34  &13.21 &--16.33$\pm$0.49  &26.05.06 &3x2700 &Flames-UVES &25\\
       &      &  &                       &                    &            &            &            &--16.75$\pm$0.38  &19.07.06 &3x2700 &Flames-UVES\\
       &      &  &                       &                    &            &            &            &--16.64$\pm$0.36  &20.07.06 &2x2700 &Flames-UVES\\
        &     &  &                       &                    &            &            &            &--16.81$\pm$0.31  &27.07.06 &2x2700 &Flames-UVES\\
(25) &        73  &2MASS 18044544-3004174  &18 04 45.44  &--30 04 17 &13.62  &13.11  &12.99 &209.43$\pm$0.34  &26.05.06 &3x2700 &Flames-UVES &30\\
       &      &  &                       &                    &            &            &            &209.53$\pm$0.57  &19.07.06 &3x2700 &Flames-UVES\\
       &      &  &                       &                    &            &            &            &209.23$\pm$0.60  &20.07.06 &2x2700 &Flames-UVES\\
       &      &  &                       &                    &            &            &            &208.95$\pm$0.42  &27.07.06 &2x2700 &Flames-UVES\\
(26) &         75 &2MASS 18045025-3004211  &18 04 50.25  &--30 04 21 &12.41  &11.76  &11.56 &112.47$\pm$0.24  &26.05.06 &3x2700 &Flames-UVES &38\\
       &      &  &                       &                    &            &            &            &112.35$\pm$0.36  &19.07.06 &3x2700 &Flames-UVES\\
       &      &  &                       &                    &            &            &            &112.57$\pm$0.35  &20.07.06 &2x2700 &Flames-UVES\\
       &      &  &                       &                    &            &            &            &112.29$\pm$0.26  &27.07.06 &2x2700 &Flames-UVES\\
(27) &         76 &2MASS 18045282-3004331  &18 04 52.81 &--30 04 33 &13.40  &12.85  &12.70  &115.82$\pm$0.24   &26.05.06 &3x2700 &Flames-UVES &25\\
       &      &  &                       &                    &            &            &            &115.58$\pm$0.49  &19.07.06 &3x2700 &Flames-UVES\\
      &       &  &                       &                    &            &            &            &115.60$\pm$0.37  &20.07.06 &2x2700 &Flames-UVES\\
      &       &  &                       &                    &            &            &            &115.49$\pm$0.31  &27.07.06 &2x2700 &Flames-UVES\\
(28) &        80  &VVV J180446.26-300200.99  &18 04 46.26  &--30 02 00 &14.05  &13.53  &13.46 &214.22$\pm$0.37  &26.05.06 &3x2700 &Flames-UVES &25\\
       &      &  &                       &                    &            &            &            &213.99$\pm$0.76  &19.07.06 &3x2700 &Flames-UVES\\
       &      &  &                       &                    &            &            &            &214.87$\pm$0.88  &20.07.06 &2x2700 &Flames-UVES\\
       &      &  &                       &                    &            &            &            &213.77$\pm$0.41  &27.07.06 &2x2700 &Flames-UVES\\
(29) &        82  &2MASS 18045450-3003063  &18 04 54.50 &--30 03 06 &13.05  &12.51  &12.37  &--106.31$\pm$0.27   &29.06.06 &2x2700 &Flames-UVES &22\\
       &      &  &                       &                    &            &            &            &--106.36$\pm$0.28  &01.07.06 &2x2700 &Flames-UVES\\
       &      &  &                       &                    &            &            &            &--105.86$\pm$0.43  &19.07.06 &2x2700 &Flames-UVES\\
(30) &        83  &2MASS 18044436-3003051  &18 04 44.36  &--30 03 05 &13.37  &12.85  &12.77 &140.23$\pm$0.43  &29.06.06 &2x2700 &Flames-UVES &49\\
        &     &  &                       &                    &            &            &            &139.66$\pm$0.49  &01.07.06 &2x2700 &Flames-UVES\\
       &      &  &                       &                    &            &            &            &140.06$\pm$0.73  &19.07.06 &2x2700 &Flames-UVES\\
(31) &        84  &2MASS 18045102-3004227  &18 04 51.02 &--30 04 22 &10.34  &9.52   &9.309   &--113.85$\pm$0.62  &29.06.06 &2x2700 &Flames-UVES &15\\
        &     &  &                       &                    &            &            &            &--113.75$\pm$0.63  &01.07.06 &2x2700 &Flames-UVES\\
      &       &  &                       &                    &            &            &            &--113.41$\pm$0.59  &19.07.06 &2x2700 &Flames-UVES\\
(32) &        86  &2MASS 18045165-3004389  &18 04 51.64  &--30 04 39 &10.68   &9.89   &9.08   &--122.46$\pm$0.30  &29.06.06 &2x2700 &Flames-UVES &16\\
       &      &  &                       &                    &            &            &            &--121.82$\pm$0.47  &31.05.06 &2x2700 &Flames-UVES\\
             \hline
(33) &   3014$^{c}$  &2MASS 18044744-3003432  &18 04 47.38  &--30 03 43 &13.49  &13.01  &12.88 &212.00$\pm$2.50 &15.08.99 &3x1200 &HIRES &34\\
(34) &   3025$^{c}$  &VVV J180447.67-300335.85  &18 04 47.62  &--30 03 36 &13.51  &12.98  &12.90 &216.80$\pm$1.50 &02.06.00 &6x1200 &HIRES &34\\
           \hline
\end{longtable}
\tablefoot{The first column provides the number that we use to identify the stars in the finding chart (see below). The names of the targets used in the fits-headers of original observations are given in the second column. This enables for a better traceability when comparing with papers that have published analyses of some of these stars. These names are also short-hand names and will be used in the text. We also list VVV and 2MASS identifications (when available) in column 3 as well as the coordinates in column 4 and 5. De-reddened magnitudes and measured radial velocity are given from column 6 to 9. Some observation informations, such as date, exposures time, and instrument, are shown from column 10 to 12. In the last column, S/N measured from the spectrum is also included. $^{a}$ De-reddened magnitudes have been derived adopting the
  reddening law by \citet{1989ApJ...345..245C}. $^{b}$ Signal-to-noise
  ratio of the final co-added spectrum. $^{c}$ The radial velocity is
  adopted from \citet{2001AJ....122.1469C}. The last column gives the
  S/N.  The average S/N of the final co-added spectrum for each star
  was estimated using the $\mathit{SPLOT}$ task within IRAF at three
  short wavelength regions ($574.4-574.7$\,nm, $604.7-606.3$\,nm, and
  $606.8-607.6$\,nm).  }
\end{tiny}
\end{longtab}

The basis for our study of stars in NGC\,6528 is our own observations
of 7 stars that were observed in 2001. In order to have a statistically
significant sample of spectra of red giant stars in NGC\,6528 we
further searched the ESO and the KECK archives. In the archive we
found spectra of 34 stars that are potential members of NGC\,6528. Of
these 12 have been observed by UVES on VLT in stand-alone mode, 20
with UVES as part of FLAMES observations, and 2 stars had been
observed with HIRES on KECK. Table\,\ref{gcs} lists basic information
about the stars and  observations.  There is no overlap between the
observations from the different programmes. Below we describe the
different data-sets in some detail.

\paragraph{Stars observed with UVES:} 
Our starting data-set consists of observations of seven stars carried
out in 2001 at the VLT 8-m telescope using the UVES spectrograph
\citep{2000SPIE.4008..534D} under our program 067.B-0382(A) (PI:
S. Feltzing). The stars were selected as members of NGC\,6528 based on
a proper motion study using high-resolution images and $VI$ photometry
from \textit{HST} \cite[see][]{2002A&A...385...67F}. In the ESO
archive we found an additional five stars observed with UVES from
programme 065.L-0340(A) and 067.D-0489(A) (PI: D. Minniti). These were
all observed in 2000 and 2001, respectively.

The resolution of the observations is 45000 or 55000 in the red arm
(480--680 nm) depending on the slit width used (0.8$"$ or 1.0$"$). The
observations were all done using the standard setup 580 with CD
\#3. The exposures have been split in order to allow for the removal
of cosmic ray hits. The data were reduced using the UVES pipeline \citep{2000Msngr.101...31B}, including bias,
inter-order background subtraction, flat field correction, extraction,
and wavelength calibration. All the echelle orders were then merged
into a 1D spectrum for each exposure.

In the archive we found 20 stars which are potential members
of NGC 6528 from program 077.B-0327(A) (PI: M. Zoccali) which was
observed in 2006.  

The multi-object optical spectrograph FLAMES
\citep{2002Msngr.110....1P} has two parts, one being FLAMES-UVES which
has 8 robotic fibers which feed the UVES spectrograph at VLT. The
spectra have $R \sim 47000$ and the observations were done with the
setup centred at 580 nm (wavelength coverage: 480--680 nm). We reduced
the raw frames using the FLAMES-UVES pipeline in a standard way. After
all the recipes, which include creating a master bias and the master
slit flat-field frame, determining the fiber order table, the
wavelength solution, and extracting the science frame, of the pipeline
have been executed we subtract the sky background spectrum from the
stellar spectra. For FLAMES-UVES the sky background is usually
observed by one fiber placed in an empty part of the sky. With the sky
background removed, we shifted the sky-subtracted spectra to a
heliocentric reference frame making use of the packages
$\mathit{RVSAO}$ and $\mathit{DOPCOR}$ within IRAF \footnote{IRAF is distributed by the
  National Optical Astronomy Observatory, which is operated by the
  Association of Universities for Research in Astronomy (AURA) under
  cooperative agreement with the National Science Fundation.}. 

To improve the S/N, we co-added
all the UVES spectra of the same star into a single spectrum. The average S/N of the final
co-added spectrum for each star was estimated by making use of the
$\mathit{SPLOT}$ task within IRAF at three
short wavelength regions ($574.4-574.7$\,nm, $604.7-606.3$\,nm, and
$606.8-607.6$\,nm). The S/N is greater than 30 per pixel for most of
the spectra (see Table~\ref{gcs}).

\paragraph{Stars observed with HIRES:}
We collected spectra of two red horizontal branch stars (3014 and
3025) from the Keck Observatory archive. The observations were taken
with HIRES \citep{1994SPIE.2198..362V} at Keck I under the programme
C53H and C19H (PI: J. Cohen) in 1999 and 2000, respectively.

Since the HIRES detector is not large enough to yield full spectral
coverage, there is a gap ($1.0-6.0$\,nm) between any two echelle
orders. The raw spectra were reduced using the MAKEE \footnote{{http://www.astro.caltech.edu/~tb/ipac\_staff/tab/makee/index.html}}
package. It determines the position of each echelle order, defines the
object and background extraction boundaries, optimally extracts a
spectrum for each order, and compute wavelength calibrations
non-interactively.

We co-added the spectra of each star into one spectrum. The measured
mean S/N of the co-added spectra was estimated in the same way as for
the UVES spectra (Table~\ref{gcs}).


\subsection{Additional data for NGC6528}

\subsubsection{Photometry}

In order to estimate good starting values for $T_{\rm{eff}}$ for our
stars we make use of colour-temperature-metallicity relations. To this
end we collected near-infrared photometric data from 2MASS
\citep{2003tmc..book.....C} and VVV (VISTA Variables in Via Lactea)
\citep{2010NewA...15..433M}.

The survey area of VVV is fully imaged in five photometric bands: $Z,
Y, J, H, K_{\rm{s}}$ and 2MASS has imaging in $J, H, K_{\rm{s}}$. VVV
goes about 4 magnitutes deeper than 2MASS. However, the larger
telescope aperture of VISTA results in saturation for the brightest
stars. The saturation magnitude in the VVV photometry occurs at
$K_{\rm{s}}<$ 12. We therefore adopt 2MASS photometry for stars 
brighter than 12 mag in $K_{\rm{s}}$ and we require that the
photometric quality flag is AAA, i.e. the best quality. For all other
stars we use VISTA photometry.

There is a very small offset between the zero-points of the VVV and
2MASS catalogs. To ensure that all photometry is on the same system we
have chosen to transfer the VVV photometry onto the 2MASS photometric
system as this is the system most commonly used in the available
temperature calibrations. The following zero-point offsets have been
applied: $(J_{\rm{2mass}}-J_{\rm{vvv}})=0.029$;
$(H_{\rm{2mass}}-H_{\rm{vvv}})=-0.039$; $(K_{\rm{s, 2mass}}-K_{\rm{s,
    vvv}})=-0.028$ \citep{2011A&A...534A...3G}.

\subsubsection{Reddening}

Reddening is important for the initial estimate of $T_{\rm eff}$, but
it can also be of interest when discussing the properties of the
cluster. Here we make use of the recent reddening maps by
\citet{2011A&A...534A...3G,2012A&A...543A..13G} to estimate the
reddening towards our stars.

As the maximum sky resolution of the reddening map is 2$^{'}$, we
assumed that the stars are subject to the same reddening within a
$4^{'} \times 4^{'}$ box. The observed targets are spread in a
relatively large area, the total coverage of the dereddened region is
$28^{'} \times 28^{'}$ consisting of 49 boxes of $4^{'} \times
4^{'}$. The mean extinction, A$_{K_{\rm{s}}}$, and colour excess,
E$(J-K_{\rm{s}})$, are calculated using
BEAM\footnote{http://mill.astro.puc.cl/BEAM/calculator.php} which uses
the extinction law from \citet{1989ApJ...345..245C}.  The adopted
coefficients for $J, H,$ and $K_{\rm{s}}$ (in the 2MASS passbands)
are: A$_{J}$ = 1.692E$(J-K_{\rm{s}})$, A$_{H}$ =
1.054E$(J-K_{\rm{s}})$, A$_{K_{\rm{s}}}$ = 0.689E$(J-K_{\rm{s}})$.
Where A$_{K_{\rm{s}}}$ = 0.689E$(B-V)$. The dereddened magnitudes of
$J, H,$ and $K_{\rm{s}}$ are listed in Table\,\ref{gcs}.

\subsubsection{Radial velocities} 
\label{sect:rv}

\begin{figure}
\begin{center}
\includegraphics[scale=0.5]{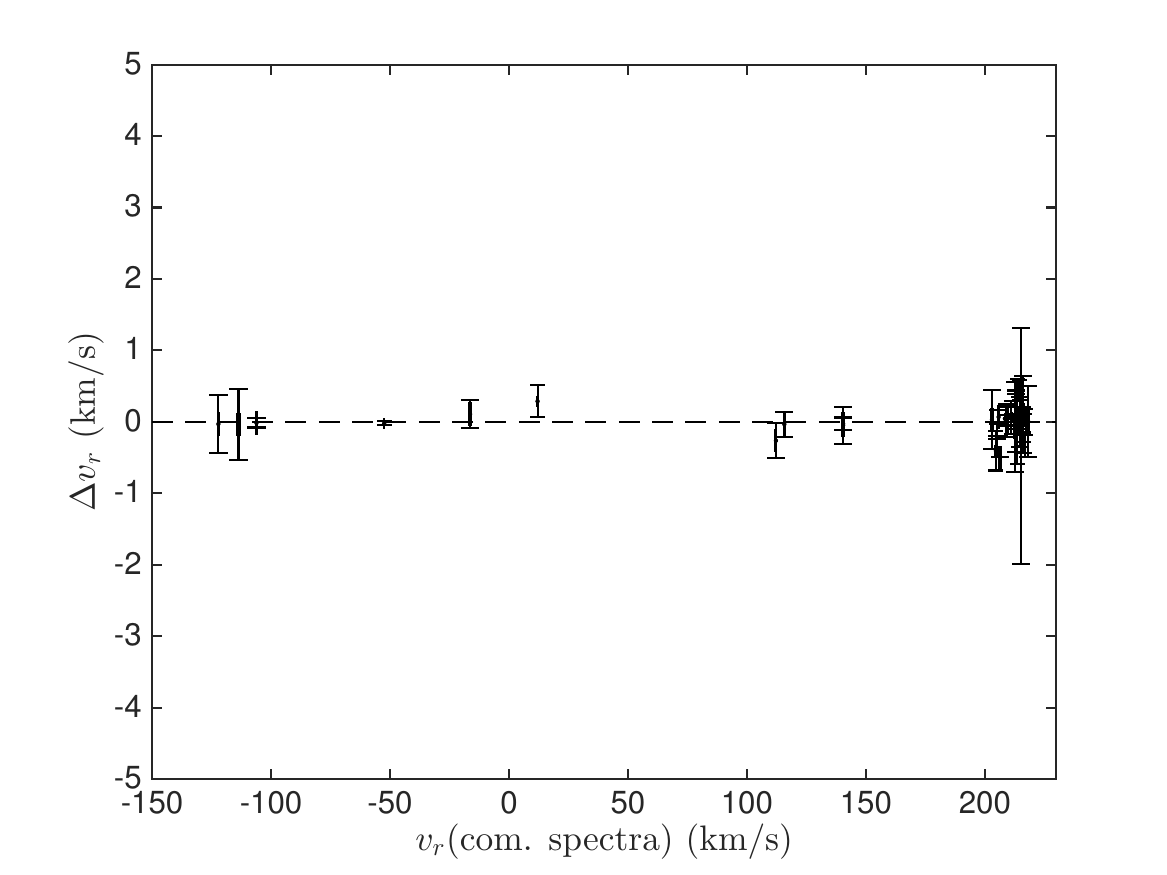}
\caption{Comparison of radial velocities for the stars derived from
  the combined spectrum and from the individual spectra separately and
  then averaged (see Sect.\,\ref{sect:rv}). The $x$-axis shows the
  value derived from the combined spectra whilst the $y$-axis shows
  the difference between $v_r$ determined in the two ways.  The
  error-bars for the difference is the coadded errors estimated from
  IRAF (for the combined spectra) and the $\sigma$ around the mean for
  the independent measures. Errors on the $x$-axis are too small to be
  seen.  }
\label{rv_comp}
\end{center}
\end{figure}

Radial velocities ($v_{\rm{r}}$) with respect to the Sun were derived
using the $\mathit{RVSAO}$ package within IRAF. The $\mathit{XCSAO}$
task cross-correlates a template spectrum with the observed spectrum
and reports the velocity difference between the two. Here we used a
template spectrum from the radial velocity standard star $\epsilon$
Peg. The resolution of the spectrum is similar to that for the spectra
of our sample spectra.

Each of our stars have more than one observation
(Table~\ref{gcs}). The spectra are often taken at different nights and
sometimes even months apart. This allows us to derive $v_{\rm{r}}$ in
two ways, enabling a test of the quality of our 
results. First we
extracted all spectra and corrected them for Earth's motion in order
to co-add them. Radial velocities were then determined using these
combined spectra. In the other approach we derived the radial velocity
for each observation individually. These individual measurements were
then averaged and the scatter around the mean was
calculated. Typically, the scatter around the mean radial velocities
is less than 0.50 km s$^{-1}$. We find that the two approaches give
very similar results (compare Fig.~\ref{rv_comp}).

For unknown reasons, we are unable to recover the radial velocities
determined by \citet{2001AJ....122.1469C} for star 3014 and 3025. Our
results are about 50 km s$^{-1}$ larger than the values given by
them. As far as we understand we are using the same spectra and same
reduction software. In Table~\ref{gcs} we list the value from
\citet{2001AJ....122.1469C}. Apart from this difference our derived
velocities agree well with those in the literature (see
Sect.\,\ref{sect:mem}).

\subsection{Membership}
\label{sect:mem}

\begin{figure}
\begin{center}
\includegraphics[scale=0.5]{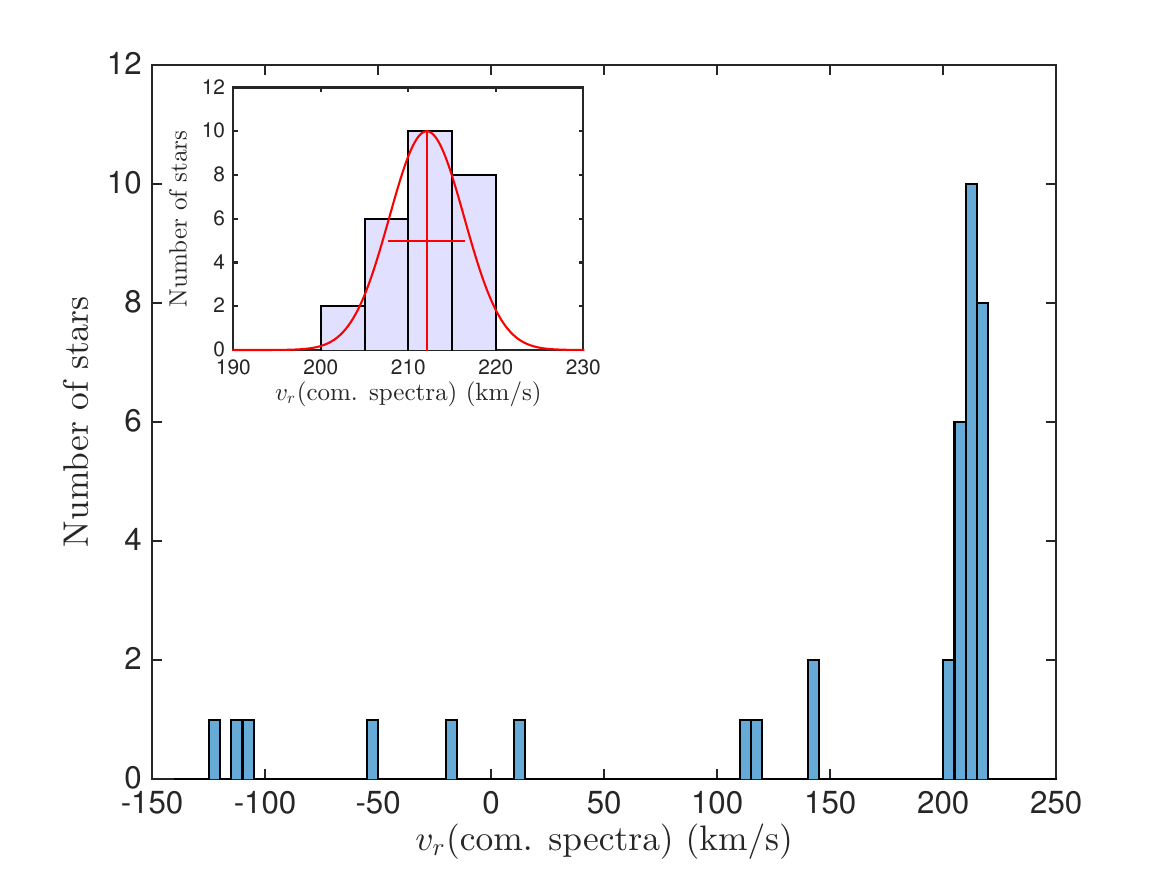}
\caption{Distribution of measured heliocentric radial velocities for
  our program stars. The inset in the top-left corner shows a zoom-in
  high-lighting the velocity distribution of the cluster members
  fitted by a Gaussian. The red cross indicates the mean (212.1 km
  s$^{-1}$) and standard deviation (4.2 km s$^{-1}$) of the radial
  velocity for the cluster members.}
\label{rvd}
\end{center}
\end{figure}

Membership of globular clusters for individual stars is commonly
determined using radial velocities \citep[e.g.,][]{2013A&A...556A..58S} or proper motions relative to the
background and foreground stars
\citep[e.g.,][]{2001AJ....121.2638Z}. NGC\,6528 has a radial velocity
of $\sim$ 210\,km\,s$^{-1}$
\citep{2001AJ....122.1469C,2004A&A...423..507Z} and a proper motion
relative to the background bulge population of $<l>$ = 0.006 and $<b>$
= 0.044 arcsec per century \citep{2002A&A...385...67F}. This means
that radial velocity is the best way to determine membership for this
cluster. Using measurements from the luminous stars in
globular clusters, as well as their integrated spectra,
\citet{1993ASPC...50..357P} determined the velocity dispersions for
many globular clusters. They found that typical velocity dispersions
in globular clusters is less than 10 km s$^{-1}$.

Figure\,\ref{rvd} shows the distribution of radial velocities for all
our stars. There is a clustering of stars just above
200\,km\,s$^{-1}$. To find the mean velocity and velocity dispersion of NGC\,6528, we
started by considering stars with a radial velocity in the range
$155 - 240$\,km\,s$^{-1}$. For these we calculated the mean and
standard deviation ($\sigma$) of the radial velocity. We then
proceeded to exclude stars more than 3$\sigma$ from the mean value and
the mean and $\sigma$ were re-calculated. This procedure was iterated
until no more stars could be excluded with a 3$\sigma$-clipping. This
left 26 stars for which we found a mean velocity of 212.1 km s$^{-1}$
and $\sigma=$ 4.2 km s$^{-1}$. Our velocity dispersion is in good
agreement with the value of 4.0 km s$^{-1}$ measured by
\citet{2001AJ....122.1469C}.

\section{Linelist and analysis tool}
\label{sect:lat}

\subsection{Line list }


To assemble the line list we made use of several sources. For iron we
selected 106 clean lines between 470 nm and 690 nm from
\citet{2013A&A...555A.150T}, \citet{2014ApJS..215...23D}, and the
linelist compiled for the {\sl Gaia}-ESO survey (Heiter et al., in prep.~and
references therein)\citep[see also,][]{2015PhyS...90e4010H}. In particular, we selected 90 \ion{Fe}{i} and 16
\ion{Fe}{ii} lines.  All our lines were examined in the spectra of the
Sun and $\xi$Hya. $\xi$Hya is a relatively cool and metal-rich star
with stellar parameters similar to those in our sample of stars in
NGC\,6528 and stars in the Galactic bulge (compare values in
Table\,\ref{sp}). Firstly, these two spectra were used to check that the lines
are clean and possible to analyse in metal-rich giant stars. 
Secondly, the equivalent width (EW) of lines in the two stars were measured. Because the strong lines are sensitive to the selection of the microturbulent velocity \citep{2006ApJ...636..821F} and might suffer from the effects of improper modeling of the outer layers in stellar atmosphere models \citep{1995AJ....109.2736M}, all lines with EWs higher than 130 m$\AA$ were removed from our line list.

As cool and metal-rich stars have more molecular lines than warmer stars, we make an extra check to make sure that our lines are not blended by molecules. For this check, we used the spectrum synthetic program (see described in \,\ref{sect:sme}) and the complete VALD \citep[][see also references in the
  data-base]{2000BaltA...9..590K, 1999A&AS..138..119K} line list containing all metal and molecular lines to compute a synthetic spectrum. To generate a spectrum of a cool and metal-rich giant, the typical stellar parameters ($T_{\rm eff} = 4500 $ K, $\log g = 2.0$, [Fe/H] = +0.1 dex) were interpolated in the MARCS model atmospheres \citep{2008A&A...486..951G}. We examined all the lines by eye and found that most of them are free of TiO, CN, C2, and MgH within 0.2 m$\AA$ of the center of the Fe line. Sixteen Fe lines might suffer a weak contamination from the CN lines. This suggests that our final measured stellar parameters and abundance should not have a large  systematic errors caused by the blending of molecular lines.

The oscillator strength ($\log gf$) for the H$\alpha$ line is adopted
from the VALD database. The atomic
data for the \ion{Ca}{i} lines are taken from
\citet{1981JPhB...14.4015S}, apart from the $\log gf$ values for
\ion{Ca}{i} 612.2 and 616.2\,nm, which are from
\citet{2009A&A...502..989A}.

The full list of lines, including references to atomic data, can be
found in the on-line Appendix\,\ref{app:linelist} (Table~\ref{loggf}).

\subsection{Spectral anlaysis  -- SME }
\label{sect:sme}

We use Spectroscopy Made Easy \citep[SME,][]{1996A&AS..118..595V,
  2005ApJS..159..141V} to determine the stellar parameters by
comparing synthetic spectra with observed spectra. SME uses the Levenberg-Marquardt (LM) algorithm to optimize stellar parameters by fitting observed spectra with synthetic spectra. The LM algorithm combines gradient search and linearization methods to determine parameter values that yield a chi-square ($\chi^{2}$) value close to the minimum. Stellar parameters and atomic line data are required to generate a synthetic spectrum. In addition to specified narrow wavelength segments of the observed spectrum, SME requires line masks in order to compare with synthetic spectrum and determine velocity shifts, and continuum masks that are used to normalize the spectral segments. The homogeneous segments and masks are created to fit all of our solar sibling candidates.

SME needs input values for $T_{\rm eff}$, $\log g$, [Fe/H], microturbulence ($v_{\rm{mic}}$), macroturbulence ($v_{\rm{mac}})$, and rotational velocity ($v{\rm \sin}i$). Given initial stellar parameters, the model atmospheres are interpolated in the precomputed MARCS model atmosphere grid, which has standard composition of elemental abundances. We then configured SME to fit each stellar spectrum by adjusting the free parameters (see Sect.\,\ref{sect:methodology}). The convergence is (likely) speedier if we start with realistic initial
values. However, it should be noted that SME is capable of converging
to the correct values also if we start from initial values that are
far off from the correct ones (Daniel Ad\'en, {\it {priv.comm.}},
tests for the {\sl Gaia}-ESO Survey).

\subsubsection{Can our line list reproduce [Fe/H] for  stars with known metallicites?}

\begin{figure}
\begin{center}
\includegraphics[scale=0.45]{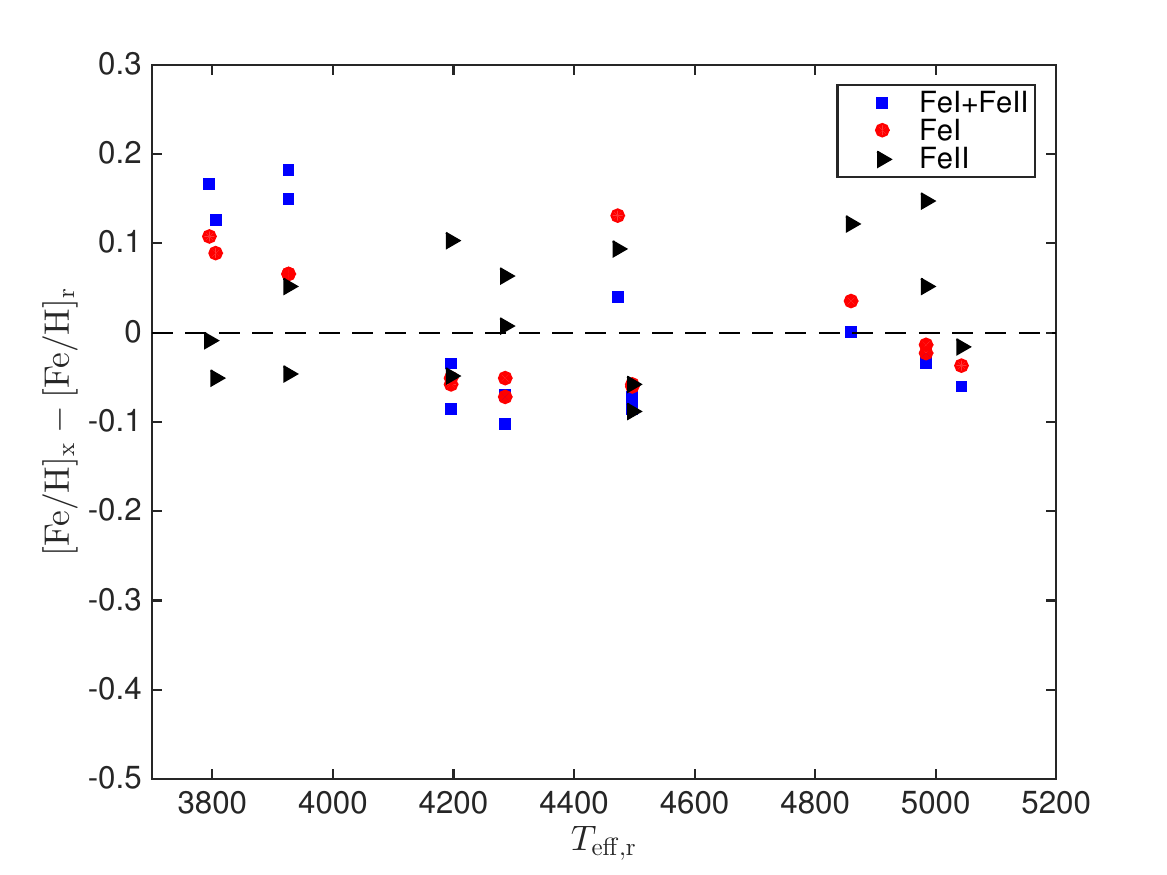}
\caption{Test of the ability of our linelist to reproduce the
  reference [Fe/H] for the {\sl Gaia} benchmark stars. The $y$-axis shows
  the difference between the value we derive, using the reference
  $T_{\rm eff}$ and $\log g$ but our own linelist and SME, and the
  reference value. We derived [Fe/H] in three ways: using both
  \ion{Fe}{i} and \ion{Fe}{ii} lines (blue filled squares), using only
  \ion{Fe}{i} lines (red filled circles), and using only \ion{Fe}{ii}
  lines (black filled triangles).  The stars can be identified using
  the $T_{\rm eff}$ values which are given on the $x$-axis. For some
  of the stars we have more than one spectrum. Results for both are shown in the plot.}
\label{metal_bm}
\end{center}
\end{figure}

Before starting our investigation, we first checked how well our set-up and
linelist reproduces the reference values for [Fe/H] for the benchmark
stars.

To this end we analysed the spectra fixing $T_{\rm
  eff}$ and $\log g$ to the recommended values (Table\,\ref{bsp}).  We
did three sets of analysis: one where both \ion{Fe}{i} and \ion{Fe}{ii}
lines were fitted, one where only \ion{Fe}{i} lines were fitted, and
one where only \ion{Fe}{ii} lines were fitted. In each case all lines
were all fitted simultaneously.

The results are shown in Fig.~\ref{metal_bm}. We find that when we fit
\ion{Fe}{i} and \ion{Fe}{ii} lines simultaneously the difference
between our values and the recommended values is 0.01\,dex with a
$\sigma$ of 0.10\,dex. Figure~\ref{metal_bm} shows that it is the
coolest stars that drives the size of the scatter, whilst the warmer stars 
have a smaller difference. When we fit only the \ion{Fe}{i}
lines the difference is 0.01\,dex with a $\sigma$ of 0.07\,dex, and
when we analyse the \ion{Fe}{ii} lines only we find a difference of
0.03\,dex with a $\sigma$ of 0.07\,dex.  There are no discernible
trends with $T_{\rm eff}$, although there is a weak indication that
\ion{Fe}{ii} lines do a better job in reproducing the reference values
for the coolest stars ($\alpha$Cet, $\gamma$Sge, and $\alpha$Tau)
while in the warmest stars ($\beta$Gem, $\epsilon$Vir, and $\xi$Hya)
\ion{Fe}{i} lines appear to do a much better better job than the
\ion{Fe}{ii} lines. However, the trend is weak and more data would be
needed to draw a firm conclusion regarding if certain species are
better at a certain temperature range for these types of stars.

This comparison shows that our line data and method of analysing the
iron lines (assuming all other parameters to be known) is fully
compatible with those in \citet{2014A&A...564A.133J} who has
established the reference values for the {\sl Gaia} benchmark stars.


\section{Looking for a robust method to determine [Fe/H] for
  metal-rich red giant stars from optical spectra}
\label{sect:methodology}

\subsection{Methods to derive effective temperature }
\label{sect:methods-teff}

Effective temperatures can be derived from stellar spectra using
several methods.  \ion{Fe}{i} is the most common ion in terms of lines
in the optical spectrum. As lines arising from the \ion{Fe}{i} ion
have a wide range of line strengths as well as excitation potentials
it is in principle straightforward to determine $T_{\rm{eff}}$ for the
star by requiring \ion{Fe}{i} lines with differing excitation
potentials to produce the same iron abundances
\citep[e.g.,][]{1993A&A...275..101E}.  This analysis rests on the
assumption of local thermal equilibrium (LTE). However, several
studies suggest that this assumption does not necessarily hold
\citep[e.g.,][]{1998A&A...338..161F,1999ApJ...521..753T,2011A&A...528A..87M,2013MNRAS.429..126R}.
Recently, NLTE calculations, e.g., from \citet{2012MNRAS.427...27B},
have been included into some studies
\citep[e.g.,][]{2013MNRAS.429..126R}.  However, at solar metallicities
the difference between LTE and NLTE results for \ion{Fe}{i} tend to be
very small for metal-rich stars \citep{2013MNRAS.429..126R} and we can
thus neglect them here.  Determination of the iron abundance is a
by-product when $T_{\rm{eff}}$ is derived in this way and errors in
the $T_{\rm{eff}}$ and iron abundance must correlate.

Another method to derive $T_{\rm{eff}}$ is to use the strong hydrogen
lines present in stellar spectra as these lines change rapidly in
strength as a function of $T_{\rm{eff}}$. These lines also have no
gravity dependence in cool stars.  Most commonly, $T_{\rm{eff}}$ is derived from the
synthesis of the extended wings of the H${\alpha}$ and H${\beta}$
lines \citep[e.g.,][]{2002A&A...385..951B,2013MNRAS.429..126R}.  This
make them very useful for determining the temperature for stars with
$T_{\rm{eff}}< 8000$\,K \citep{2005oasp.book.....G}. With this method
iron must be determined independently and hence the errors are not
correlated.

Another approach is to use line-depth ratios to derive $T_{\rm{eff}}$
\citep{1994PASP..106.1248G}. One advantage of using line-depth ratios
is that these do not depend on the metallicity of the star. In the visible spectral range, unblended Si, Ti, V, Cr, Fe and Ni line-depth rations can be used to estimate $T_{\rm{eff}}$. There are
not many applications of this method in the literature, but
\citet{2000A&A...358..587K} illustrated that the line-depth ratios are
powerful indicators of $T_{\rm{eff}}$ at least for supergiants. We
will not further explore this method here.

Finally, we can use calibration of stellar photometry to
derive $T_{\rm{eff}}$ from a colour index, such as $(B-V)_{\rm 0}$
\citep[see, e.g.,][]{2009A&A...497..497G}. This method has its
limitations. For example, for stars in the Galactic bulge or in the
plane of the Galaxy the photometric approach is hampered by the poorly
determined reddening of individual stellar colours. Although this
method is not very dependable for bulge stars, thanks to the large
reddening along the line-of-sight, it nevertheless provides useful
starting values for our analysis.

\subsection{Methods to derive surface gravity}
\label{sect:methods-logg}

It is common to measure $\log g$ by imposing ionization equilibrium
which means that we require \ion{Fe}{i} and \ion{Fe}{ii} lines to
produce the same abundance\footnote{In principle other elements can
  also be used (e.g., Ti). However, there are only a few elements that
  have lines arising from both the neutral and singly ionised variety
  of the atom that are observable in the optical part of the spectrum.
  In those cases, often one of the ions has very few or very weak
  lines.}.  \ion{Fe}{ii} lines can be weak and they are not as
numerous as the \ion{Fe}{i} lines. This method thus requires high
quality spectra with large wavelength coverage. As discussed above
(Sect.\,\ref{sect:methods-teff}) \ion{Fe}{i} lines can be susceptible to
departures from LTE. This is not true for \ion{Fe}{ii} lines. Which
means that ionizational balance can be influenced by departures from LTE.
Determination of the iron abundance is a by-product of this method and
errors in $\log g$ and [Fe/H] must correlate.

Several lines show strongly pressure-broadened wings in the spectra of
cooler stars. This means that the surface gravity of late-type stars
can be determined from analysis of these lines. Lines that have
pressure broadened wings include the \ion{Mg}{i}~b lines at 516.7,
517.2, 518.3\,nm, the \ion{Na}{i}~D lines at 588.9 and 589.5\,nm, and
the \ion{Ca}{i} lines at 612.2, 616.2, and 643.9\,nm \citep[see,
  e.g.,][]{1988A&A...190..148E, 1993MNRAS.264..334B,
  1997A&A...323..909F}. The analysis of the wings of these strong
lines require that the elemental abundance of the element the line is
arising from is also determined, from other lines than the one used to
derive $\log g$. This means that errors in $\log g$ and the abundance
of that element correlate.

Finally, when we have independent measurements of $T_{\rm eff}$ and
metallicity, $\log g$ can be derived through isochrone fitting
\citep[see, e.g.,][]{2007ApJ...664.1190S}. We will not further explore
this method here.

\subsection{Determining stellar parameters  }

In Sects.\,\ref{sect:methods-teff} and \ref{sect:methods-logg} we
discussed various ways to determine $T_{\rm eff}$ and $\log
g$. Although this is not an exhaustive summary, combining the
various ways to derive $T_{\rm eff}$ and $\log g$ still leads to quite
a few possible combinations. Here we will explore the following
combinations:

\begin{description}

\item[Method 1:] In this method we only use \ion{Fe}{i} and
  \ion{Fe}{ii} lines to constrain all stellar parameters (i.e.,
  excitation and ionizational equilibria are imposed).

\item[Method 2:] In this method we use the H$\alpha$ line to constrain
  $T_{\rm eff}$, the strong \ion{Ca}{i} lines to constrain $\log g$,
  and \ion{Fe}{i} lines to derive the iron abundance.

\item[Method 3:] In this method we use the H$\alpha$ line to constrain
  $T_{\rm eff}$, the strong \ion{Ca}{i} lines to constrain $\log g$,
  and \ion{Fe}{ii} lines to derive the iron abundance.

\end{description}

\noindent
While in Method 1 all parameters are correlated, Method 2 and 3 aim at
breaking these degeneracies by using separate measures for each of the
three main parameters. We did not attempt to use the \ion{Mg}{i} b and
\ion{Na}{i} D lines as these are too blended by the atomic and
molecular lines to determine $\log g$ from them in metal-rich giant
stars as we are interested in here. For cool giants with $T_{\rm{eff}}
<$ 4400 K, we further found that the wings of the H$\alpha$ line
almost vanished (see also Appendix\,\ref{app:Halpha}). This lead us to
modify Method 2 to be able to include also such stars in our study
(see Sect.\,\ref{sect:method2}).

In addition to $T_{\rm eff}$, $\log g$, and [Fe/H] a spectroscopic
analysis normally calls for the determination of a few other
parameters; microturbulance ($v_{mic}$), marcotublence ($v_{mac}$),
and rotational velocity ($v{\rm \sin}i$).

Microturbulance and macroturbulence 
can either be derived from the spectrum itself or
obtained from a calibration, such as those  in
\citet{2014A&A...564A.133J}. Rotational velocities of red giants and
horizontal branch stars are typically small ($v{\rm \sin}i < 5$ km
s$^{-1}$), therefore a good assumption is to set $v{\rm \sin}i$ to 1
km s$^{-1}$. Which is what we do for the rest of this paper. We note
that there is often a degeneracy between $v{\rm \sin}i$ and $v_{mac}$
which is hard to break in the analysis. For the type of study presented
here -- where the aim is to get the best measure of [Fe/H] -- the 
exact values of these two parameters is of less relevance to the final
outcome of the investigation and we can allow them to be degenerate. This
implies that in the this study we can regard them (together) as nuisance
necessary to fit but not necessarily giving physical insight in to the 
star. 

To test the three methods we adopted ten benchmark stars
\citep[see][and
  Sect.\,\ref{sect:BS}]{2015arXiv150606095H,2014A&A...564A.133J}.

\subsubsection{Method 1: \ion{Fe}{i} and \ion{Fe}{ii} lines  }
\label{sect:method1}

In this method we fit all free parameters simultaneously by
comparing the observed spectrum with a synthetic one for short regions
around the \ion{Fe}{i} and \ion{Fe}{ii} lines. It should be mentioned that the method is more commonly performed with equivalent widths in the literatures \citep[e.g.,][]{2003A&A...410..527B}. This method is speedier if we start with realistic initial values for
stellar parameters (see also
Sect.\,\ref{sect:sme}). For the benchmark stars, their recommended
values are used as the initial input, however, experience shows that
the exact values are unimportant to achive convergence.  An initial
value for $v_{\rm mic}$ was obtained using the relation given in
\citet{2014A&A...564A.133J}.  The initial $v_{\rm mac}$ value was set
to 5.0 km~s$^{-1}$ for all stars.

All parameters are simultaneously fitted inside SME. As SME strives to fit all lines equally
well this is in effect the same as requiring ionizational and
excitational equilibrium -- hence, all parameters are determined
simultaneously. The method differs slightly from its most common
implementation in the literature where an iterative scheme is employed
and one parameters is varied in each step \citep[see, e.g.,][and
  discussions therein]{1991MNRAS.251..369D,1991MNRAS.250...89D}. Instead, the usage of SME
in fully free mode is more akin to the method used in
\citet{2009A&A...493..913F} where the parameter space is searched (by
hand) for a best fit to all criteria simultaneously.

\subsubsection{Method 2: H$\alpha$, Ca, and \ion{Fe}{i} lines}
\label{sect:method2}

\begin{figure*}
\begin{center}
\includegraphics[scale=0.4]{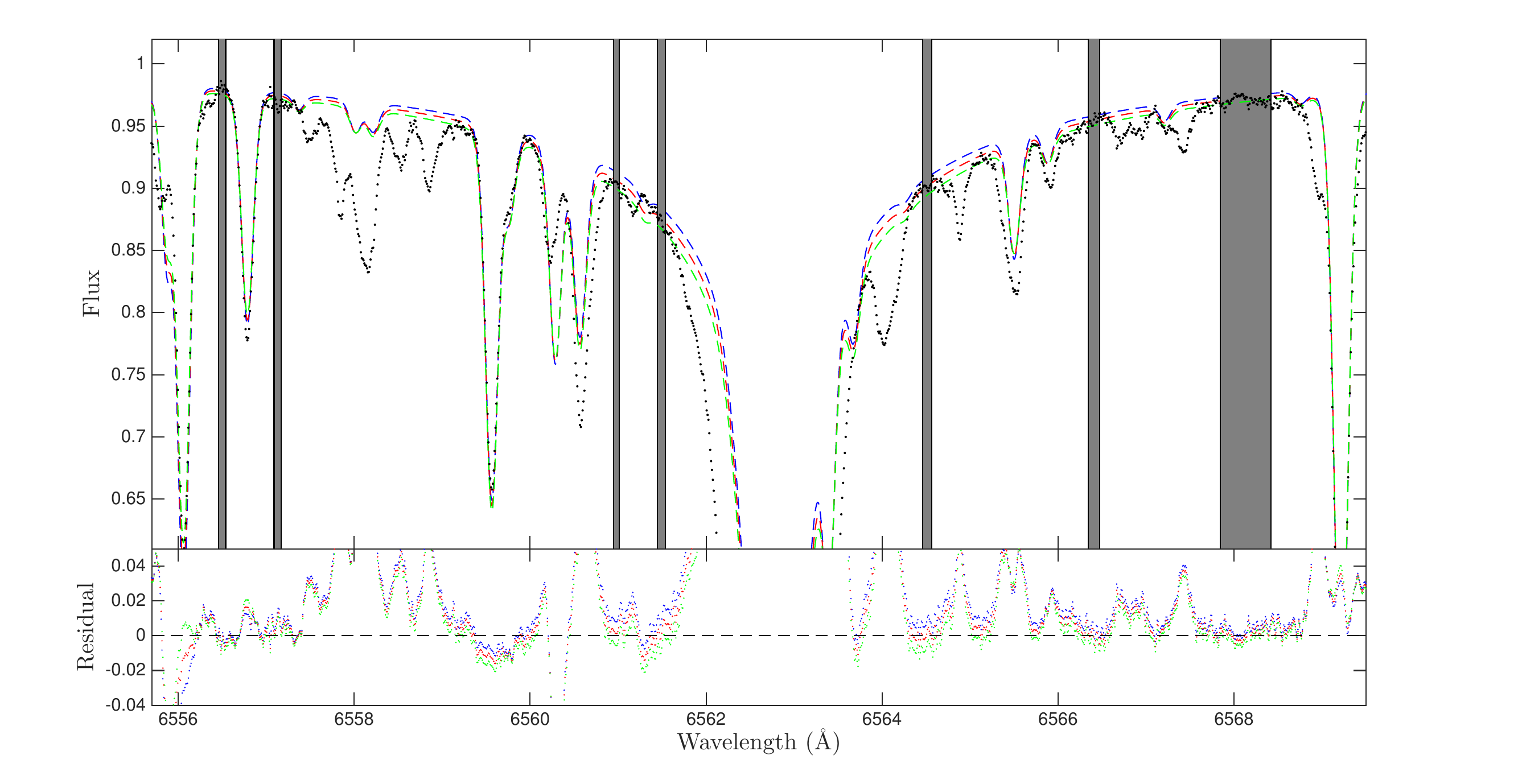}
\caption{Comparison of synthetic spectra with the wings of H$\alpha$
  line in $\epsilon$Vir. The middle synthetic spectrum (in red) shows
  the best fit. The other two synthetic spectra (blue and green)
  indicate the shape of the wings when $T_{\rm{eff}}$ is changed
  according to our estimated uncertainty (here 50\,K). The gray areas
  mark the regions used to evaluate the goodness of the fit. Note that
  these are relatively narrow since the spectrum has very little clean
  "line continuum" thanks to the cool temperature of the star, which
  results in many lines being present in the spectrum.}
\label{vir}
\end{center}
\end{figure*}

\begin{table*}
\centering
\setlength{\tabcolsep}{6pt}
\begin{tiny}
\caption{Stellar parameters for the {\sl Gaia} benchmark stars. Columns 2 to
  4 give the reference values for the stellar parameters. 
  Columns 5 till 11 list our results, derived using Method 2 and
  Modified Method 2 (Sect.\,\ref{sect:method2}).}
\label{bsp}
\begin{tabular}{l|rrr|rrrrrrrrr}
      \hline
      \hline
       & \multicolumn{3}{c|}{Reference values} &  \multicolumn{7}{c}{Results from Method 2} \\
      Name  &$T_{\rm{eff,r}}$ &$\log g_{\rm{r}}$ &[Fe/H]$_{\rm{r}}$ &$T_{\rm{eff}}$ &$\sigma_{T_{\rm{eff}}}$&$\log g$  &[Fe/H] &$\sigma_{\rm{[Fe/H]}}$ &$V_{\rm{mic}}$ &$V_{\rm{mac}}$\\
                   &(K) && (dex)&(K) &(K) && (dex)  & (dex)&(km~s$^{-1}$) &(km~s$^{-1}$)\\
      \hline
$\mu$Leo      &4474 &2.51 &  0.25 &4513  &100  &2.50   &  0.31  &0.08   &1.5     &4.8\\
HD107328      &4496 &2.09 &--0.33 &4484  &50   &2.00   &--0.48  &0.09   &1.8     &4.4\\
              &---  &---  &---    &4483  &50   &2.01   &--0.45  &0.06   &1.8     &5.4\\
$\beta$Gem    &4858 &2.90 &  0.13 &4846  &50   &2.89   &  0.11  &0.08   &1.2     &4.2\\
$\epsilon$Vir &4983 &2.77 &  0.15 &5056  &50   &2.84   &  0.16  &0.08   &1.5     &5.5\\
              &---  &---  &---    &5055  &50   &2.80   &  0.17  &0.08   &1.5     &5.7\\
$\xi$Hya      &5044 &2.87 &  0.16 &4991  &50   &2.96   &  0.10  &0.08   &1.4     &6.1\\
\hline
       & \multicolumn{3}{c|}{} & \\
       & \multicolumn{3}{c|}{} &  \multicolumn{7}{c}{Results from Modified Method 2 } \\
\hline
Arcturus      &4286 &1.64 &--0.52 &4305  &100   &1.69   &--0.58  &0.09   &1.7     &5.4\\
              &---  &---  &---    &4321  &100   &1.63   &--0.61  &0.09   &1.7     &5.2\\
$\alpha$Cet   &3796 &0.68 &--0.45 &3888  &100   &0.72   &--0.43  &0.12   &1.6     &8.9\\
$\gamma$Sge   &3807 &1.05 &--0.17 &3988  &100   &0.96   &--0.17  &0.09   &1.6     &6.6\\
$\alpha$Tau   &3927 &1.11 &--0.37 &3910  &100   &1.12   &--0.27  &0.11   &1.4     &5.7\\
              &---  &---  &---    &3930  &100   &1.10   &--0.28  &0.10   &1.4     &6.1\\
$\beta$Ara    &4197 &1.05 &--0.05 &4250  &100   &1.03   &--0.17  &0.08   &3.0    &10.4\\
              &---  &---  &---    &4277  &100   &1.26   &--0.13  &0.08   &3.0     &9.8\\
       \hline
           \hline
\end{tabular}
\tablefoot{The first column gives the name of the star. The second and
  third columns give the recommended effective temperature and surface
  gravity obtained from \citet{2015arXiv150606095H}, while the
  recommended metallicity from \citet{2014A&A...564A.133J} is listed
  in the column 4. For some stars we have two spectra from different
  instruments (see details in Sect.\,\ref{sect:BS}). The second spectrum is
  indicated by a dash in these three columns. The following seven
  columns give our results.}
\end{tiny}
\end{table*}

To break the degeneracy between the different parameters we use
different spectral indicators for each of the main stellar
parameters. This method requires a set of initial parameters to get
started. These could be obtained in different ways (e.g., from
photometry). For this test we take the parameters obtained in Method 1
as initial input.

\smallskip

\noindent
{\sl Method 2} :
For  warm giants ($T_{\rm{eff}} > 4400$ K), the following steps are used:
\begin{enumerate}
\item A set of parameters as initial input;
  
\item Fit the wings of H$\alpha$ to derive $T_{\rm{eff}}$, while the
  other parameters are kept fixed;
  
\item Fit the wings of the three strong \ion{Ca}{i} lines ($\lambda$
  612.2, 616.2, and 643.9\,nm) to determine $\log g$. To break the
  degeneracy of $\log g$ and Ca abundance, several weak \ion{Ca}{i}
  lines are also fitted at the same time;
  
\item Derive [Fe/H] from the \ion{Fe}{i} lines by setting both
  $v_{\rm{mic}}$ and $v_{\rm{mac}}$ free, while the other parameters
  are kept fixed at the values derived in the steps above;
  
\item Repeat steps 2 to 4 with updated parameters until all five
  parameters reach convergence.
\end{enumerate}

For step 2 a number of synthetic spectra are generated and their fit
is evaluated using the spectral regions indicated in
Fig.\,\ref{vir}. Further examples can be found in the on-line
material in Appendix\,\ref{app:Halpha}.

In Step 3 the three strong \ion{Ca}{i} lines are simultaneously
fitted. It would be possible to fit them each individually instead. A
discussion of the merits of the different approaches can be found in
the on-line material in Appendix\,\ref{app:logg}.

Step 5 require us to decide when a given parameter has converged. For the
current work the following convergence criteria were applied: between
two iterations the differences should be
$\Delta T_{\rm{eff}} < 50$\,K for most stars, but for stars with very high 
$S/N$-ratios we required $< 20$\,K, $\Delta \log g < 0.05$\,dex, and 
$\Delta$[Fe/H]$< 0.05$\,dex. 

\medskip

As mentioned, it is not feasible to measure $T_{\rm eff}$ from the
H$\alpha$ line for the cooler giants ($T_{\rm{eff}} < 4400$
K). Therefore, the only option is to derive $T_{\rm eff}$ using
\ion{Fe}{i} lines (excitational equilibrium) rather than the wings of
the H$\alpha$ line for these stars. However, this re-introduces a
degeneracy between $T_{\rm eff}$ and [Fe/H]. We are still able to use
the three strong \ion{Ca}{i} lines to determine $\log g$.

\smallskip
\noindent
{\sl Modified Method 2} :
For  cool giants ($T_{\rm{eff}} < 4400$\,{K},  the following steps are used:
\begin{enumerate}
\item A set of parameters as initial input;
\item Fit the wings of the three strong \ion{Ca}{i} lines ($\lambda$
  612.22, 616.22, and 643.91\,nm) to determine $\log g$. To break the
  degeneracy of $\log g$ and Ca abundance, several weak \ion{Ca}{i}
  lines are also fitted at the same time;
\item Setting both $v_{\rm{mic}}$ and $v_{\rm{mac}}$ free while
  keeping $\log g$ fixed, we only fit \ion{Fe}{i} lines to determine
  $T_{\rm{eff}}$ and [Fe/H] at the same time;
\item Repeat steps 2 to 3 with updated parameters until all five
  parameters converge.
\end{enumerate}

\subsubsection{Method 3: H$\alpha$, Ca, and \ion{Fe}{ii} lines}
\label{sect:method3}

Since the only difference between Method 3 and Method 2 is that we use
\ion{Fe}{ii} rather than \ion{Fe}{i} lines, the atmospheric parameters
are measured making use of the same iterative process described in
Sect.\,\ref{sect:method2} by changing from \ion{Fe}{i} line to
\ion{Fe}{ii} lines.

For Modified Method 2 we did make use of excitation equilibrium using
the \ion{Fe}{i} lines. However, the number of \ion{Fe}{ii} lines
available is much fewer than \ion{Fe}{i} lines and they do not cover a
large enough range in excitation potential for us to try to determine
$T_{\rm eff}$ from excitation equilibrium.  Method 3 is thus only
implemented for warm giants ($T_{\rm{eff}}> 4400$\,K).

\subsection{Discussion and error estimates}
\label{sect:discussion}

\begin{figure*}
\includegraphics[trim = 30mm 10mm 30mm 5mm, clip, width=18cm]{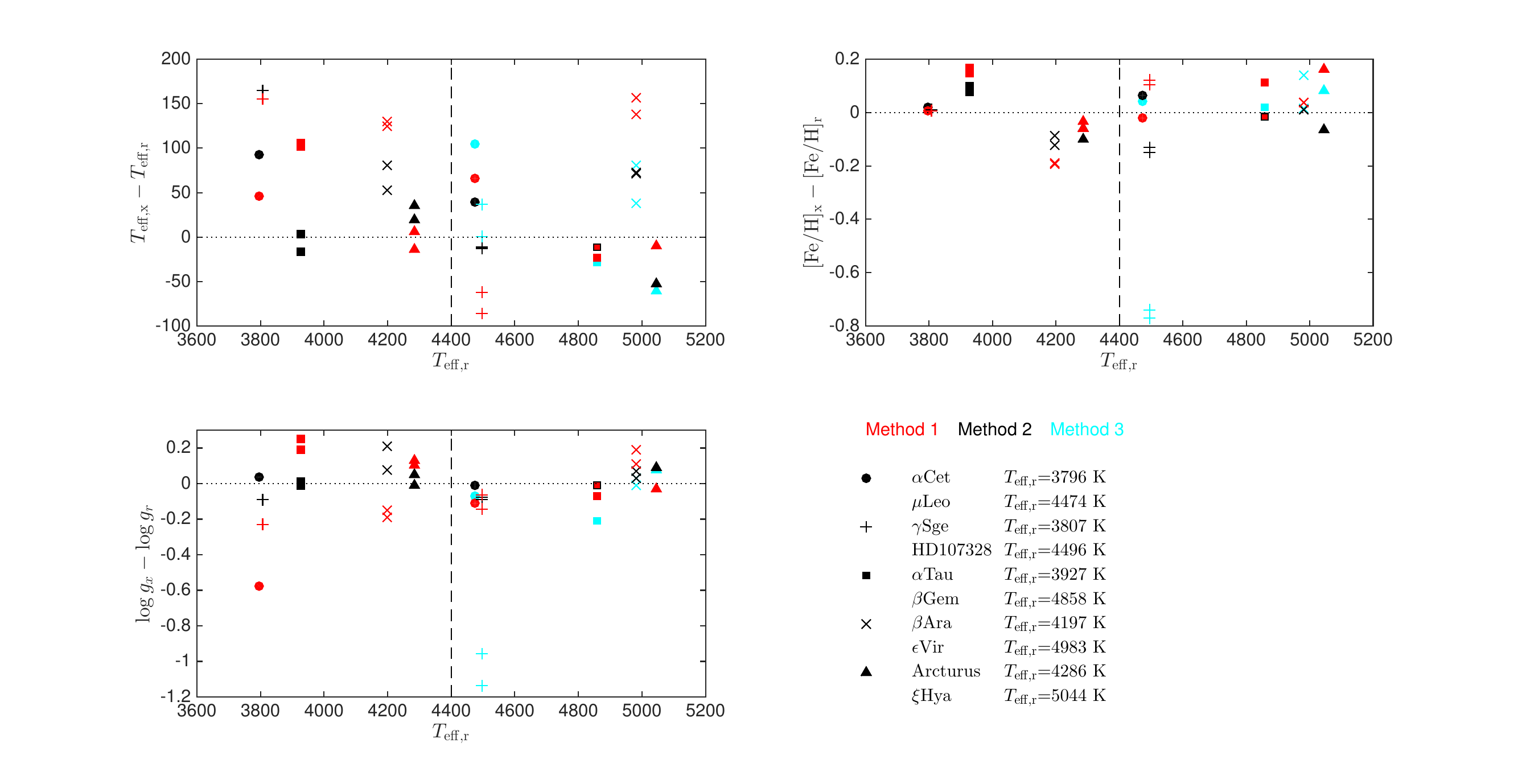}
\caption{Comparison of the stellar parameters derived using the three
  methods discussed in Sects.\,\ref{sect:method1} till
  \ref{sect:method3} with the reference values (Table\,\ref{bsp}). On
  each $y-$-axis is indicated the difference between the values we
  have obtained and the reference value (indicated by a subscript
  "r"). Method 1 is indicated by red symbols, Method 2 by black, and
  Method 3 is indicated by cyan symbols. Different symbols represent
  the different stars (as indicated in the figure, note that each
  symbol represent two stars.They should be easy to identify thanks to
  their different $T_{\rm eff} $ ). Note that some stars have more
  than one spectrum analysed (see Sect.\,\ref{sect:BS}). The stars can be
  identified by their recommended $T_{\rm eff} $ which is plotted on
  the $x$-axis.The dashed line indicates the separation of "cool"
  (left) and "warm" (right) giants (see Sect.\,\ref{sect:method2}).}
\label{fig:BSresults}
\end{figure*}

We now proceed to compare our results for the ten benchmark stars
derived using the three methods. A comparison is shown in
Fig.~\ref{fig:BSresults} and the stellar parameters obtained using
Method 2 and Modified Method 2 are listed Table~\ref{bsp}.  In
particulare we find \footnote{For these
  estimates we excluded the star HD\,107328 as the [Fe/H] for that
  star is very different, and we do not have a straightforward answer
  to this discrepancy.}:

\begin{description}
\item{\bf Method 1} $\Delta T_{\rm{eff}}$ = 56\,K ($\sigma$=79),
  $\Delta \log g$ = --0.04 ($\sigma$=0.20), and $\Delta$[Fe/H] =
  0.03\,dex ($\sigma$=0.11);

\vspace{0.2cm}
\item{\bf Method 2} $\Delta T_{\rm{eff}}$ = 35\,K ($\sigma$=54),
  $\Delta \log g$ = 0.02 ($\sigma$=0.07), and $\Delta$[Fe/H] = --0.03\,dex
  ($\sigma$=0.08);

  \vspace{0.2cm}
\item{\bf Method 3} $\Delta T_{\rm{eff}}$ = 19\,K ($\sigma$=53),
  $\Delta \log g$ = --0.03 ($\sigma$=0.08), and $\Delta$[Fe/H] =
  0.04\,dex ($\sigma$=0.04) ({\bf warm giants only}).

\end{description}

Of the three methods method 1 clearly performs the least well. It is
noticeable that the scatter in all three stellar parameters are quite
a bit higher for this method than for the other two. The offsets are
also somewhat larger. From Fig.~\ref{fig:BSresults} we see that there
is no particular trend in $\Delta T_{\rm{eff}}$, $\Delta \log g$, and
$\Delta$[Fe/H] as a function of $T_{\rm eff}$. This is reassuring as
it indicates that our methods do not have hidden trends in them and
that the estimated scatters can be trusted.

Method 3 suffers from the shortcoming that it can not be applied to
cooler giants ($T_{\rm{eff}} < 4400$ K). This is a severe limitation
for the study of globular clusters and field giants in the Galactic
bulge. To study the metallicity distributions, we would therefore
recommend Method 2 as providing a sound basis for an analysis of a
large sample of stars. It is reassuring that this method not only
gives good metallicity estimates but also reliable stellar parameters
in general. 

The uncertainty in $T_{\rm{eff}}$ can be estimated by inspecting the
comparison of synthetic spectra with the observed spectrum. This is
relatively straightforward for the high S/N and high resolution
spectra we have for the {\sl Gaia} benchmark stars. We found that the
typical uncertainty in $T_{\rm{eff}}$ is about 50 K for most of the
warm giants. One example, $\epsilon$Vir, is shown in Fig.~\ref{vir}. For $\log g$, we found that for most stars the typical uncertainty is around 0.10 dex (see more details in Appendix\,\ref{app:logg}).

For [Fe/H] we calculate the error in the mean by using the iron abundance from each iron line
  indiviudally (remember that the parameters are decoupled in this
  method). Calculating the formal error in the mean we find errors
  typically around 0.02 dex, whilst $\sigma$ is just below 0.1 dex
  (see Table~\ref{bsp}).

For the benchmark stars, the typical uncertainties in the recommended
$T_{\rm{eff}}$ and $\log g$ are about 60 K and 0.10 dex \citep{2015arXiv150606095H}, respectively.
The scatter in $T_{\rm{eff}}$ and $\log g$ around the recommended
values for the benchmark stars found in this study is comparable with the
typical uncertainties of the recommended stellar parameters. There is
a very small offset between the recommended [Fe/H] and our results
when using Method 2. The typical uncertainty in the recommended [Fe/H]
is comparable to what we find \citep[$<$ 0.02
  dex,][]{2014A&A...564A.133J}.

\subsection{How much do the results depend on the S/N in the spectra?}
\label{sect:snrsb}

Our tests of different methods have relied on the analysis of high S/N
spectra of the metal-rich, cool giants in {\sl Gaia} benchmark
sample. Clearly, a study of for example the Galactic bulge will have
spectra of much lower S/N. How will this influence the results? We
analyse this in two steps. Firstly we will degrade the spectra for the
{\sl Gaia} benchmark stars to mimic that of high quality data for stars in
the Galactic bulge and secondly we will make use of the extensive data
set of stars in NGC\,6528 that we have collected.

To test the dependence of our final results on S/N and resolution
we have  degraded
the spectra for the ten benchmark stars and re-analysed them.  We chose 45\,000 and 20\,000 for the
resolution and at three different S/N (25, 35, 45). 20\,000 is a
resolution that is common for FLAMES-GIRAFFE (used in the {\sl Gaia}-ESO
Survey) and the future massively multiplex spectrographs \citep[e.g.,
  WEAVE and
  4MOST,][respectively]{2014SPIE.9147E..0LD,2014SPIE.9147E..0MD}.

Comparing with the results derived from the original spectra we found
that [Fe/H] is weakly overestimated when the spectra are noisier, but
[Fe/H] does not depend on the resolution of the spectrum. It should be
noted that we did not investigate if the abundances of other elements
were unaffected. The iron lines we use are carefully selected
to be the cleanest and most easy to analyse, for other elements this
may not be true.


\section{Applying our analysis to stars in NGC\,6528}
\label{sect:6528}

\subsection{Stellar parameters for NGC\,6528 stars}\label{sect:sp}

Initial $T_{\rm eff}$ for the stars were calculated using the
colour-$T_{\rm{eff}}$-[Fe/H] relation from
\citet{2009A&A...497..497G}. As we could not obtain photometric data
in the optical for all our stars, the dereddened $J-K_{\rm{s}}$ colour
is used to calculate the initial estimate of $T_{\rm{eff}}$. Following
previous studies of the metallicity of NGC\,6528
\citep[e.g.,][]{2004A&A...423..507Z}, we set [Fe/H] = --0.10 dex as
the initial value for the analysis for all stars. Since most of our
stars are either horizontal branch or red clump stars, we simply set
the initial $\log g$ to 2.5\,dex for warm giants ($T_{\rm{eff}} \ge$
4400 K) and 2.0\,dex for the cooler. For the analysis we used Method 2
and Modified Method 2, as desribed in Sect.\,\ref{sect:method2}.

Inspecting the spectrum of star 1-24, we found that most of the lines
are asymmetric. \citet{2004A&A...423..507Z} found that this star
likely is a binary. Our inspection of the spectrum supports this
conclusion. The star was removed from our sample.

For stars 61, 66, 68, and 84 we were unable to analyse the spectra as
they all have  low S/N and in addition the spectra appear to have
several unexpected features. These stars were excluded from the
analysis.

This leaves us with 29 stars successfully analysed. The results are
listed in Table\,\ref{sp}. Eight of these stars are not radial velocity
members of NGC\,6528 (see Sect.\,\ref{sect:mem}).

\begin{table*}
\centering
\setlength{\tabcolsep}{1pt}
\caption{Stellar parameters for the sample stars derived using Method 2.}
\label{sp}
\begin{tabular}{lccrrrrrrrrrr}
      \hline
      \hline
 \#  &   Object  &$T_{\rm{eff}}$ &$\sigma_{T_{\rm{eff}}}$&$\log g$ &$\sigma_{\log g}$ &[Fe/H] &$\sigma_{\rm{[Fe/H]}}$ &$V_{\rm{mic}}$ &$V_{\rm{mac}}$ &Member\\
 &     &(K) &(K) && &  & &(km~s$^{-1}$) &(km~s$^{-1}$)\\
      \hline
(01) &01   &4550  &100 &2.78  &0.02 &0.40  &0.12 &1.8    &4.2  &N\\
(02) &02   &6042  &50 &4.51  &0.05 &0.03  &0.04 &1.1    &8.2  &N\\
(03) &03   &4029  &100 &1.23  &0.15 &--0.15  &0.12 &1.5    &5.3 &Y\\
(04) &04   &4304  &100 &1.52  &0.15 &0.04  &0.11 &1.4    &5.3  &Y\\
(05) &05   &4277  &100 &1.95  &0.11 &0.01  &0.10 &1.5    &4.5  &Y\\ 
(06) &06   &4623  &100 &1.98  &0.08 &--0.04  &0.14 &1.9    &6.8  &Y\\
(07)  &07   &4776  &100 &2.50  &0.09 &0.03  &0.08 &1.7    &6.1  &Y\\ 
(08) &1-02   &4258  &100 &1.67  &0.12 &0.05  &0.09 &1.3    &9.8  &Y\\
(09) &1-16   &4640  &100 &2.56  &0.14 &--0.02  &0.09 &1.5    &7.1  &Y\\
(11) &1-36   &4337  &100 &2.14  &0.08 &--0.08  &0.11 &1.7    &4.8  &Y\\
(12) &1-42   &4125  &100 &1.39  &0.19 &0.09 &0.14 &0.8    &7.8  &Y\\
(13) &60   &4289  &100 &2.08  &0.09 &--0.07  &0.08 &1.7    &6.2  &Y\\
(15) &62   &3967  &100 &1.89  &0.10 &--0.08  &0.11 &1.8    &7.1  &Y\\ 
(16) &63   &3977  &100 &1.45  &0.07 &--0.15  &0.09 &1.8    &7.0  &Y\\
 (17) &64   &4079  &100 &1.94  &0.15 &--0.07  &0.12 &1.7    &5.6  &Y\\
(18) &65   &4289  &100 &2.28  &0.08 &--0.09  &0.12 &1.9    &6.5  &Y\\ 
(21) &69   &4044  &100 &1.65  &0.08 &--0.17  &0.12 &1.7    &6.2  &Y\\
(22) &70   &4123  &100 &2.12  &0.12 &0.09  &0.20 &1.1    &7.3  &Y\\
(23) &71   &4560  &150 &2.64  &0.04 &0.11  &0.08 &1.9    &7.7  &Y\\ 
(24) &72   &4506  &150 &2.55  &0.04 &0.26  &0.13 &2.2    &6.7  &N\\
(25) &73   &4702  &100 &2.62  &0.20 &--0.02  &0.10 &2.4   &10.9  &Y\\
(26) &75   &4153  &100 &2.00  &0.13 &0.01  &0.10 &1.7    &4.8  &N\\
(27) &76   &4569  &150 &2.74  &0.22 &0.43  &0.13 &2.1    &5.4  &N\\
(28) &80   &4451  &100 &2.51  &0.05 &0.19  &0.14 &1.7    &6.4  &Y\\
(29) &82   &4521  &100 &2.34  &0.17 &0.40  &0.12 &2.0    &5.2  &N\\
(30) &83   &4883  &100 &2.48  &0.21 &--0.48  &0.07 &1.7    &5.6  &N\\
(32) &86   &3650  &150 &1.15  &0.38 &0.03  &0.34 &1.7   &10.0  &N\\
(33) &3014   &4797  &100 &2.21  &0.10 &0.21  &0.11 &1.4    &6.7  &Y\\
(34) &3025   &4877  &100 &2.02  &0.15 &0.14  &0.09 &1.7   &15.8  &Y\\
      \hline
           \hline
\end{tabular}
\tablefoot{Column one gives the  \#  from Table\,\ref{gcs}  in Appendix\,\ref{app:FC}, which are also used in the Finding Chart in the same appendix. Column two the original IDs, Columns three to eight lists the derived parameters and their associated errors.  Columns nine and ten list the micro and macro turbulences, respectively. "Y" and "N" in the last column indicate if the star is a
  radial velocity member of NGC\,6528 or not.}
\end{table*}

\subsection{Analysis of a suite of low S/N spectra for stars in NGC\,6528}
\label{sect:SNR6528}

\begin{figure}
\begin{center}
\includegraphics[scale=0.45]{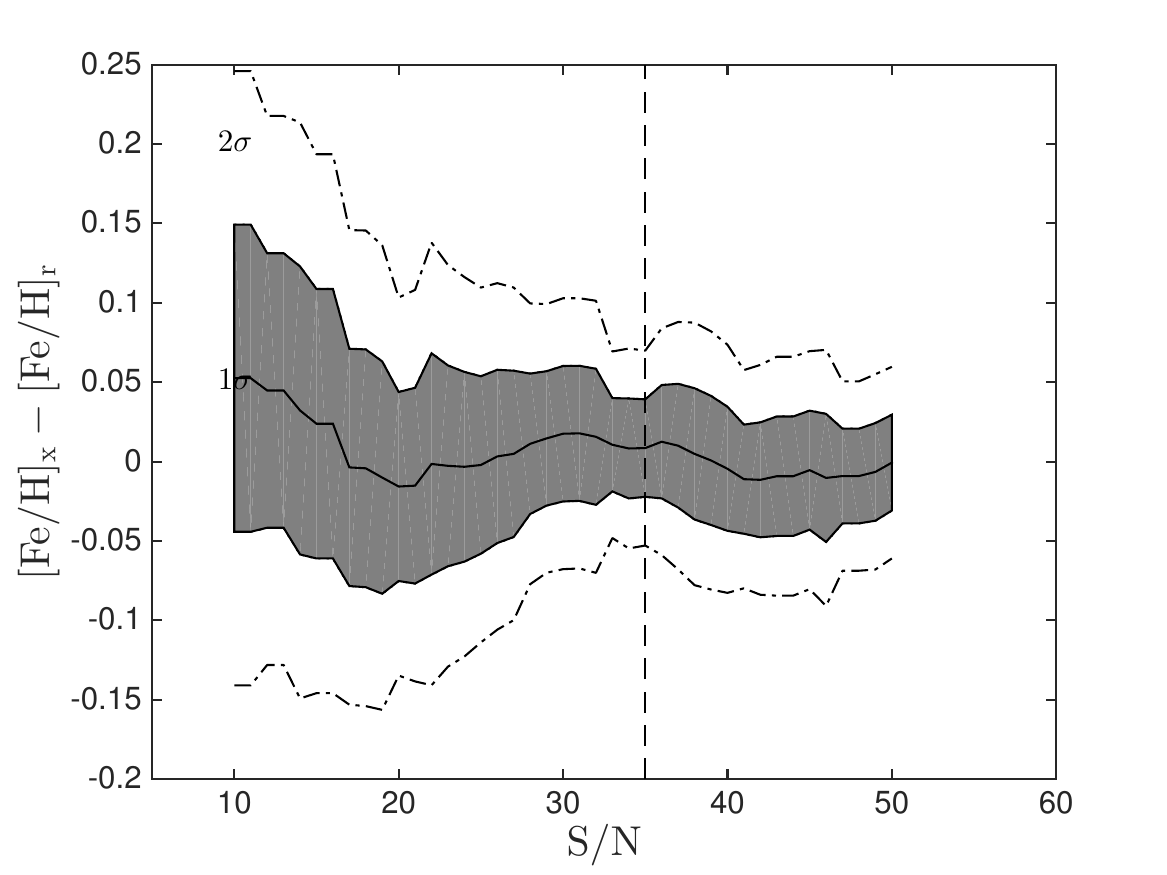}
\caption{The difference between [Fe/H] derived from spectra of varying
  S/N (per pixel) for stars in NGC\,6528.  [Fe/H]$_{\rm{x}}$ indicates our value
  measured for each spectrum, while [Fe/H]$_{\rm{r}}$ indicate the
  value obtained in Sect.~\ref{sect:sp} for each star. The full line
  indicated the average difference evaluated as a running average. The
  shaded region indicates the 1\,$\sigma$ difference, while the
  dashed-dotted line indicates the 2\,$\sigma$ difference. Details on
  individual stars are given in online material in
  Appendix\,\ref{app:SNR}. Note that in this figure we do not
  distinguish between stars for which different methods to derive
  $T_{\rm eff}$ have been used (compare Fig.\ref{fig:fehcum}).}
\label{msnr}
\end{center}
\end{figure}

For several of the stars in NGC\,6528 we have multiple exposures. This
allows us to analyse spectra of different S/N for the same star.
Using Method 2 and Modified Method 2 we analysed the actual observed
spectra of five stars (star-05, -07, -62, -65 and -67, IDs as in
Table\,\ref{gcs}).

Two of the cluster stars (-62 and -71) have a large number of spectra
allowing us to study the results in a more statistical way.
For reference value we use the final parameters determined for each
star as listed in Table\,\ref{sp}.

We find that the scatter, for a given star and for the whole sample,
increases as we go to spectra with lower and lower S/N. This is not a
new finding, but it is here quantified in detail for, what we believe,
is the first time for evolved, metal-rich giant stars. Although the
scatter increases as we go to lower S/N the mean value of the
measurements more or less reproduces the reference value.  At the
lowest S/N we find a positive offset (Fig.\,\ref{msnr}).

Further details are available in the online material in Appendix\,\ref{app:SNR}.

\subsection{Discussion: Do Method 2 and Modified Method 2 produce comparable results?}
\label{sect:disc}

\begin{figure*}
  \sidecaption
\includegraphics[ width=12cm]{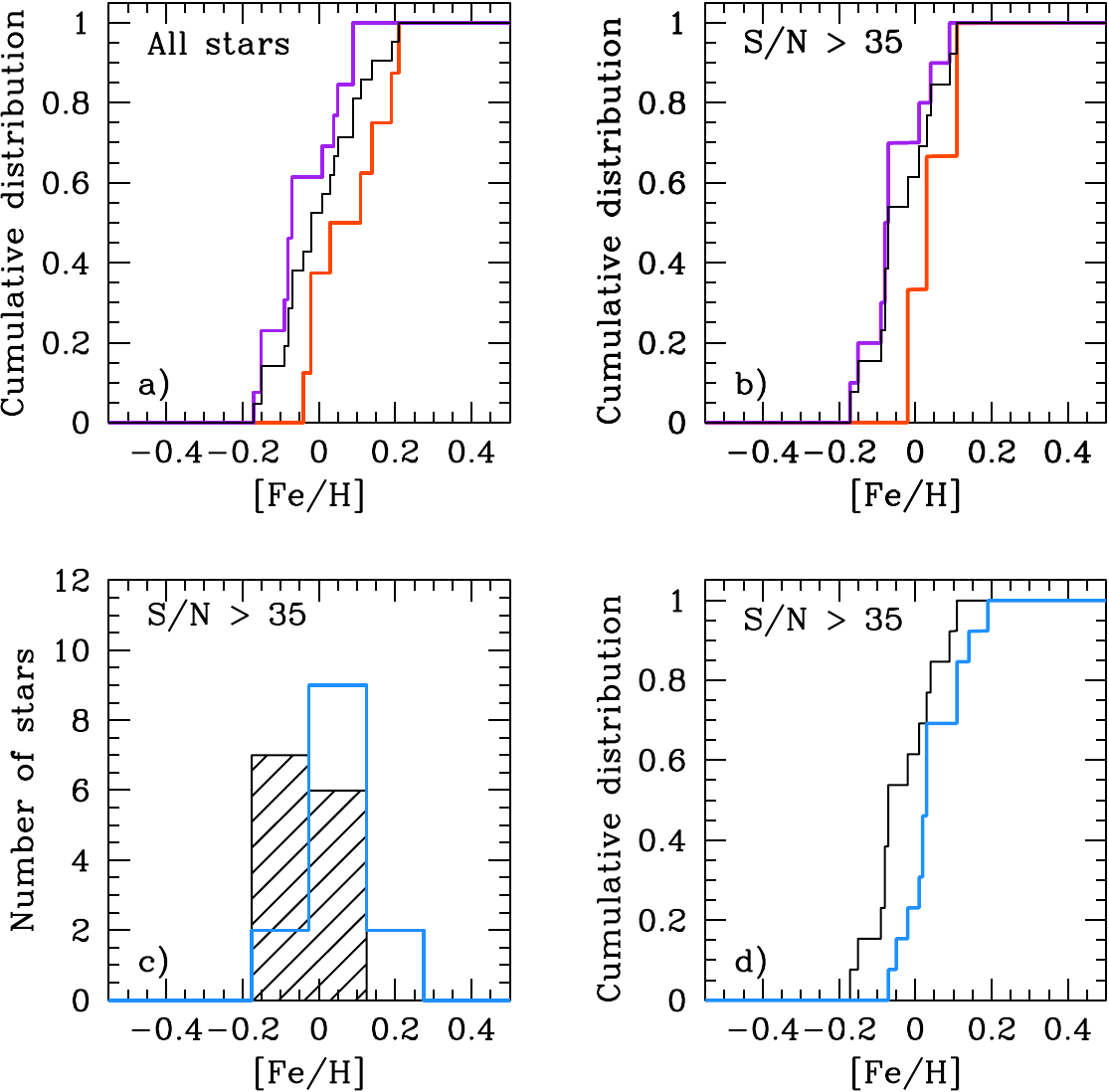}
\caption{Distribution of [Fe/H] for stars in NGC\,6528. {\bf a)} The
  black line shows the cummulative distribution of [Fe/H] for the full
  sample (21 stars). The orange-red line shows the distribution for
  stars where we have fitted the wings of Halpha to determine $T_{\rm
    eff}$ (Method 2), while the purple line shows the distribution for
  the stars when we have used excitation equilibrium to determine
  $T_{\rm eff}$ (Modified Method 2). {\bf b)} The same as in a) but
  now only for stars with S/N $>35$. {\bf c)} The black line shows the
  histogram of [Fe/H] for stars in NGC\,6528 for all stars with S/N
  $>35$, while the blue line shows the result when the offset between
  the distributions for stars analysed with Method 2 and Modified
  Method 2 has been accounted for.  {\bf d)} The cummulative
  distribution of [Fe/H]. The colours are the same as in c).  }
\label{fig:fehcum}
\end{figure*}

\begin{table}
\setlength{\tabcolsep}{7pt}
\begin{tiny}
  \caption{Metallicity estimates for NGC\,6528 using different samples.
  The two first lines in the table gibes the results using Method 2 and
  Modified Method 2 together to analyse the whole sample.
  The line marked recalibrated presents our final result. Below this,
  we give the results if the sample is split according to which method
  is used to analyse the spectra.}
\label{tab:feh6528}
\begin{tabular}{lrrrrr}
      \hline
      \hline
 & S/N& \multicolumn{1}{c}{$N_{\rm stars}$}  & \multicolumn{1}{c}{$<$[Fe/H]$>$} & \multicolumn{1}{c}{$\sigma$} & Error \\
      \hline
                  & All     & 21 &   0.00 & 0.10 & 0.02 \\
                  & $>35$   & 13 & --0.03 & 0.08 & 0.02  \\
Recalibrated      & $>35$   & 13 &   0.04 & 0.07 & 0.02  \\   
\hline
  \\
\hline
Method 2          & All     & 8  &   0.07 & 0.10 & 0.03 \\
		  & $>35$   & 3  &   0.04 & 0.06 & 0.03 \\
Modified Method 2 & All     & 13 & --0.04 & 0.09 & 0.02 \\
                  & $>35$   & 10 & --0.05 & 0.08 & 0.02 \\
\hline
\end{tabular}
\tablefoot{The first column indicates which methods are included in
  the average, the second column indicates if all S/N are included or
  if there is a cut-off, the third column gives the number of stars in
  the sample, the fifth column lists the average [Fe/H], while columns
  six and seven lists the associated $\sigma$ and error-in-the-mean
  ($\sigma/\sqrt N_{\rm stars}$). }
\end{tiny}
\end{table}

Unfortunately, the answer to the question posed in the headline of
this section is -- no. This should perhaps not come as a surprise,
but given the results in Sect.\,\ref{sect:discussion} we would have
expected that the results would be relatively similar and no (obvious)
systematic differences present. This is, however, not the case.

Figure\,\ref{fig:fehcum} shows the cumulative distributions of [Fe/H]
for stars that are radial velocity members of NGC\,6528. We plot the
distribution for all stars, but also for stars analysed using Method 2
and Modified Method 2, respectively. Table\,\ref{tab:feh6528} lists
the number of stars in each sub-sample and the average [Fe/H]. Stars
analysed using Method 2 are about 0.1\,dex more metal-rich than stars
analysed with Modified Method 2. The median is similarly off-set. To
check that this result does not depend on S/N we also analysed only
stars with S/N $>35$. Although this results in rather small samples
the end-result  remains (Fig.\,\ref{fig:fehcum} b).

We also note that when selecting only stars with S/N $>35$ the full
sample as well as the two sub-samples all give a lower [Fe/H] (by 0.03
dex). This agrees with what we found in the analysis of the noise injected
spectra of {\sl Gaia} benchmark stars (see Sects.\,\ref{sect:snrsb} and \ref{sect:SNR6528}).

To summarize, Fig.~\ref{msnr} indicate that the methods are very good
at reproducing the final [Fe/H] also at relatively low S/N, but below
about $\sim 35$ the errors increase and results in inflated values for
[Fe/H]. However, of greater concern is that the results from the two
methods are offset (Fig.\,\ref{fig:fehcum} a \& b).


\section{The metallicity of NGC\,6528}\label{sect:metallicity} 

\subsection{Results}

Given the results found in Sect.\,\ref{sect:SNR6528} and
\ref{sect:disc} defining the iron abundance of the metal-rich globular
cluster NGC\,6528 requires some care. In Method 2 the degeneracy
between $T_{\rm{eff}}$ and [Fe/H] is broken and is  thus the most robust
method to derive [Fe/H] with. Our final result for
NGC\,6528 will therefore be based on this method. In
Fig.\,\ref{fig:fehcum} (panels c and d) we show the final result when
the [Fe/H] from Modified Method 2 are calibrated to Method 2 by
applying the offset between the two methods to stars analysed with
Modified Method 2.

The final sample of stars in NGC\,6528 thus contains 3 $+$ 10 stars
with S/N $>$ 35 (compare Table\,\ref{tab:feh6528}).  If we consider
these 13 stars which have S/N $>$ 35 the mean and standard deviation
change to $+0.04$\,dex with $\sigma=0.07$ dex and hence an error in the
derived mean abundance of 0.02 dex. This is our best estimate of 
[Fe/H] for NGC\,6528.


\subsection{Comparison with literature}

The spectra of four red horizontal brach stars in NGC 6528 were
analyzed by \citet{2001AJ....122.1469C}. They found that the mean
[Fe/H] for NGC 6528 is 0.07$\pm$0.01 dex. Two of these stars (3014 and
3025) are part of our sample. For these stars our [Fe/H] are higher
than found by \citet{2001AJ....122.1469C}. The difference is
particularly large for 3014 for which our value is 0.1 dex
higher.  The discrepancy between the two results is mainly caused by
the difference in $T_{\rm{eff}}$. Our $T_{\rm{eff}}$ is hotter than
that of \citet{2001AJ....122.1469C} by about 150 K, while the $\log
g$s are similar.

Four of our stars (1-16, 1-24, 1-36, and 1-42) were taken from the
sample by \citet{2004A&A...423..507Z}. As discussed above 1-24 is a
spectroscopic binary and has been excluded from further
analysis. \citet{2004A&A...423..507Z} note that 1-16 has a nearby
companion. They therefore discarded also this star. As we have a large
sample we are less worried about slight contamination of an individual
spectrum and have kept the star in the sample.

The [Fe/H] determined by us and \citet{2004A&A...423..507Z} for 1-36
agree very well. On the other hand, the differences for I-42 are large.  Although
the two $T_{\rm{eff}}$ are very similar there is a difference in the
metallicity and gravity of about 0.20 dex. Several possibilities might
explain this discrepancy. The most obvious one being that we have
different line lists and potentially also different $\log gf$ values
for lines in comment \footnote{We can not make a comparison for the
  two list of iron lines, because \citeauthor{2004A&A...423..507Z} did
  not publish their iron line data.}. We have also made use of the
latest updates of $\log gf$-values. We found that the uncertainty in
the determined metallicity and gravity is much larger than the typical
values, because 1-42 is a cooler giant ($T_{\rm{eff}} \sim$ 4100
K). \citeauthor{2004A&A...423..507Z} also found that 1-42 has large
scatter in iron abundances. We also note that 1-42 has a lower S/N
than the other stars from \citet{2004A&A...423..507Z}. Given that it is also a cool giant and hence analysed
with Modified Method 2, it is likely that the results are quite
influenced by the deterioration in results below S/N $\sim$ 35 (see also
plots in Appendix\,\ref{app:SNR}). Thus there does not seem to be one
single issue that have contributed to the difference but several.

Apart from these two high-resolution studies, several other recent
studies are summarized in Table~\ref{mref}. Most of these agree with
our results, with the exception of the metallicity derived from a comparison 
of the low resolution spectra to a grid of synthetic spectra \citep{2001A&A...376..136C}.

\begin{table}
\setlength{\tabcolsep}{7pt}
\begin{tiny}
\caption{Metallicity estimates for NGC\,6528 from the literature.}
\label{mref}
\begin{tabular}{lrrr}
      \hline
      \hline
Method & [Fe/H] &Ref.\\
               & dex    & \\
      \hline
IR abs. at 1.6 $\mu$m &--0.23 & 1\\
Low resolution spectra &--0.50$\pm$0.3 &2\\
High resolution spectra (RHBs) &0.07$\pm$0.01 & 3 \\
RGB morphology indicators  &$\sim$--0.2 &4 \\
High resolution spectra (RGs) &--0.10$\pm$0.20  & 5 \\
High resolution infrared spectra &--0.17$\pm$0.01  & 6 \\
Photometric calibration &--0.04$\pm$0.27 &7 \\
           \hline
\end{tabular}
\tablefoot{The numbers in the last column indicate the reference: 1) \citet{1997A&A...321..859O}; 2) \citet{2001A&A...376..136C}; 3)
  \citet{2001AJ....122.1469C}; 4) \citet{2003A&A...402..607M}; 5) \citet{2004A&A...423..507Z}; 6) \citet{2005MNRAS.363..897O}; 7) \citet{2014A&A...565A...8C} }
\end{tiny}
\end{table}


\section{Conclusions and implications for the study of metal-rich stellar populations}\label{sect:conclusions}

With the aim of providing robust measures of [Fe/H]   for metal-rich red
giant branch stars we have conducted a thorough study of how best to
analyse such spectra. We first analysed spectra for a set of so called
benchmark stars to see which methods could best reproduce the
reference values and secondly we analysed a set of spectra for a
metal-rich globular cluster to appraise how well the chosen method
fared in a real example with sometimes spectra of poor quality. Our
conclusions are the following

\begin{itemize}
\item Low S/N in the spectra skews [Fe/H]  values to higher values.
  
\item Using excitation equilibrium to determine $T_{\rm eff}$ results
  in [Fe/H] that are 0.1 dex lower than if we fit the wings of the
  $H \alpha$ line are used to derive $T_{\rm eff}$.
\end{itemize}

These two effects should be taken into account when studying red giant stars in
metal-rich stellar populations. In general the best approach would be to ensure
large enough S/N in the spectra and only use stars with $T_{\rm eff} > 4400$\,K.
However, this might not always be practical or even possible. When cool stars
need to be used or when the S/N on average can not be sufficiently high the
study should implement a careful approach already at the telescope; a substantial
number of warmer stars should be observed at sufficiently high S/N. These stars
need not be part of the science sample, but can, for example, be stars in a globular
cluster or in the field. In fact, stars in a globular cluster might be preferable
as only there will it be straightforward to do the type of comparison that we
present in Sect.\,\ref{sect:disc}

We study the [Fe/H] in the metal-rich and old globular cluster NGC 6528. We find that 
the best value, taking the various issues summarised above into account, is 
[Fe/H]  = $+0.04$\,dex with $\sigma=0.07$ dex and hence an error in the
derived mean abundance of 0.02 dex.


\begin{acknowledgements}
  We thank Luca Sbordone who helped us to degrade the original {\sl Gaia}
  benchmark star spectra to lower resolution and lower S/N. These
  degraded spectra were used in our tests.
  
  This project was supported by the grant No. 621-2011-5042 from The
  Swedish Research Council. G.R. is funded by the project grant "The
  New Milky Way" from the Knut and Alice Wallenberg Foundation. This
  work has made use of the VALD database, operated at Uppsala
  University, the Institute of Astronomy RAS in Moscow, and the
  University of Vienna. This research made use of the SIMBAD database,
  operated at the CDS, Strasbourg, France.
\end{acknowledgements}

\bibliographystyle{aa} 
\bibliography{cheng-referen} 

\Online

\begin{appendix}


\section{NGC\,6528 -- a finding chart}
\label{app:FC}

\begin{figure*}
\begin{center}
\includegraphics[scale=0.75]{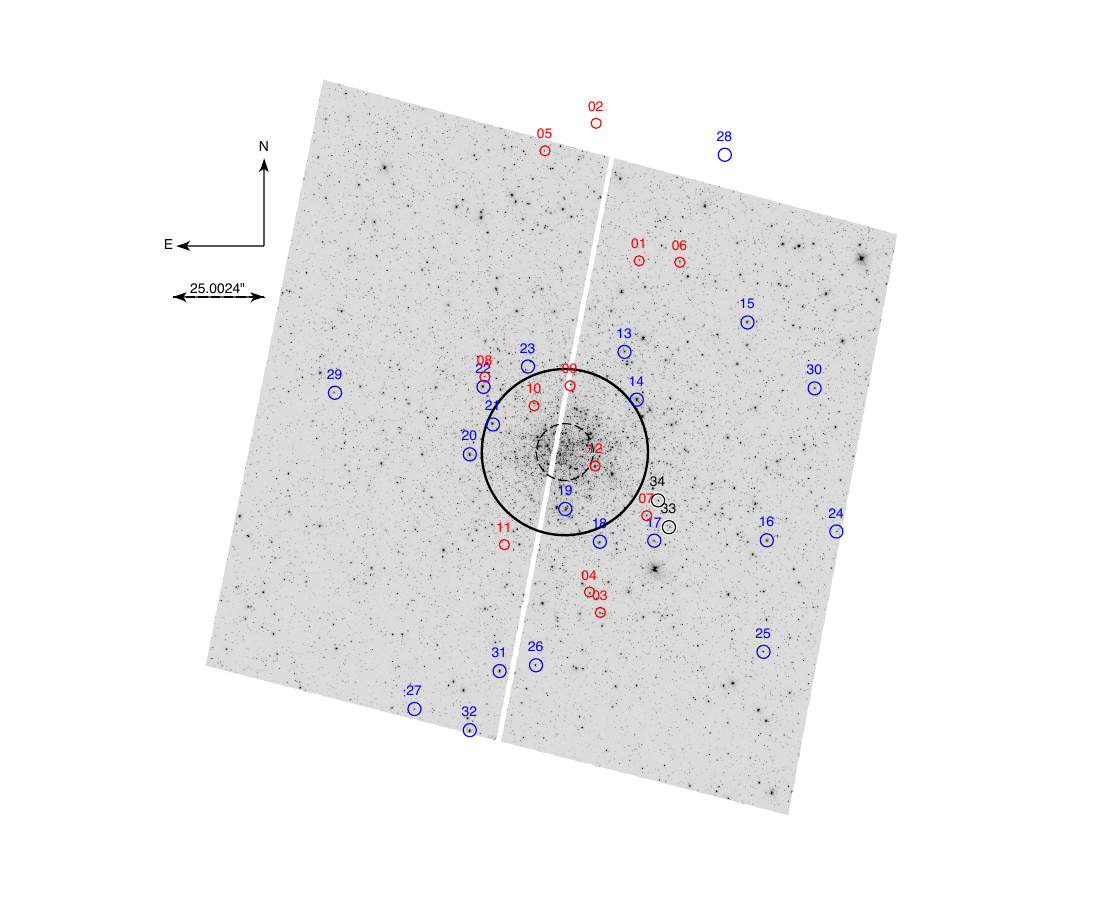}
\caption{An image of the globular cluster NGC\,6528 with the positions
  of stars we have analysed in this study marked. The stars are identified by
  the numbers listed in the first column in Table\,\ref{gcs}.  Red circles show
  stars observed with UVES, blue circles show stars observed with
  FLAMES-UVES, and black circles show stars observed with HIRES. The
  numbers next to the circles refer to the numbers in
  Table\,\ref{gcs}. The core radius of the cluster (0.13 arcmin) is
  indicated by a red dashed line and the half-light (0.38 arcmin) by a
  black solid line. Data taken from \citet{1996AJ....112.1487H}. The
  image has been smoothed with a Gaussian to roughly mimic the seeing
  at Paranal.}
\label{figapp:FC}
\end{center}
\end{figure*}

Figure\,\ref{figapp:FC} shows a finding chart centred on the field of
the globular cluster NGC\,6528. All stars from Table\,\ref{gcs} are
marked (two stars fall outside the image) and identified by the number that is listed in the first column in Table\,\ref{gcs}. The image is based on WFC3
observations with the HST and has been retrieved from the Mikulski Archive for Space Telescopes \footnote{http://archive.stsci.edu/}.


\section{Linelist}
\label{app:linelist}

This section presents the full linelist used in the study of the ten
{\sl Gaia} benchmark stars and the red giant branch stars in NGC\,6528. The
data is shown in Table\,\ref{loggf}, where also references to the
original sources of the data are given. Most of the lines have been
included in the linelist used in the {\sl Gaia}-ESO Survey
\citep[][and Heiter et al., in prep.]{2012Msngr.147...25G}. 

In the fourth column in
Table\,\ref{loggf} we give the classification used in the {\sl Gaia}-ESO
linelist. Two flags are given, the left refers to the quality of the
log {\it gf} and the second to the blending properties of the line.

\bigskip

\noindent
$\bullet$   The flags have the following meaning for log {\it gf}:
\begin{itemize}
\item  y --  Data come from a trusted source (mainly laboratory measurements with excellent accuracy).

\item  u --  Quality of data is not decided (advanced theoretical calculations and lower accuracy laboratory data).
\item  n --  Data are expected to have low accuracy.
\end{itemize}

\noindent
$\bullet$ The flags have the following meaning for blending properties as assessed in spectra of the Sun and Arcturus:

\begin{itemize}
\item y --  Line is particularly un-blended or only blended with line of same species in both stars.
\item u --  Line may be inappropriate in at least one of the stars.
\item n -- Line is strongly blended with line(s) of different species in both the Sun and Arcturus.
\end{itemize}

\onecolumn

\begin{longtab}
\setlength{\tabcolsep}{11pt}
\begin{longtable}{cccrr}
\caption{\label{loggf} Atomic line data.}\\
\hline\hline
$\lambda$    &$\chi$    &$\log gf$    & Flag & Ref.\\
                   (${\rm \mathring{A}}$)    &(eV)       \\
      \hline  \endfirsthead
      \caption{continued.}\\
      \hline\hline
$\lambda$    &$\chi$    &$\log gf$    & Flag & Ref.\\
                   (${\rm \mathring{A}}$)    &(eV)       \\
      \hline
      \endhead
      \hline
      \endfoot     
H I        &		  &		&	  &	\\
6562.797   &    10.1990   &    0.710    &         &   (4)\\
	   &		  &		&	  &	 \\
Ca I	   &		  &		&	  &	 \\
6122.217   &      1.8860  &    -0.380   &     yy  &   (1) \\
6161.297   &      2.5230  &    -1.266   &     yn  &   (2) \\
6162.173   &      1.8990  &    -0.170   &         &     (1)\\
6166.439   &      2.5210  &    -1.142   &     yy  &   (2)  \\
6169.042   &      2.5230  &    -0.797   &     yy  &   (2)  \\
6169.563   &      2.5260  &    -0.478   &     yy  &   (2)  \\
6439.075   &      2.5260  &     0.390   &     yy  &   (2)  \\
6455.598   &      2.5230  &    -1.320   &     yy  &   (2)  \\
           &              &             &         &	\\
Fe I	   &		  &		&	  &	 \\
4787.827   &     2.9980   &    -2.620   &      yu &   (3)\\
4788.757   &     3.2370   &    -1.763   &      yu &   (4)  \\
4802.880   &     3.6420   &    -1.514   &      yy &    (5) \\
4808.148   &     3.2510   &    -2.630   &      yy &   (6) \\
4843.143   &     3.3970   &    -1.650   &         &   (3)\\
4882.143   &     3.4170   &    -1.480   &      nu &   (3)\\
4910.017   &     3.3970   &    -1.280   &         &   (3)\\
4950.106   &     3.4170   &    -1.500   &      uy &   (3)\\
4962.572   &     4.1780   &    -1.182   &      yu &   (7) \\
4994.129   &     0.9150   &    -3.002   &      yy &   (7) \\
5141.739   &     2.4240   &    -2.125   &      yy &   (6)\\
5223.183   &     3.6350   &    -2.252   &      uy &   (6)\\
5228.376   &     4.2200   &    -1.095   &      uu &   (6)\\
5242.491   &     3.6340   &    -0.967   &      yy &   (7) \\
5243.776   &     4.2560   &    -1.022   &      uu &   (6)\\
5247.050   &     0.0870   &    -4.975   &      yu &   (7) \\
5253.462   &     3.2830   &    -1.580   &      yu &   (3)\\
5295.312   &     4.4150   &    -1.518   &      uy &   (6)\\
5322.041   &     2.2790   &    -2.802   &      yy &   (7) \\
5379.574   &     3.6940   &    -1.514   &      yy &   (7) \\
5386.333   &     4.1540   &    -1.670   &      uy &   (7) \\
5389.479   &     4.4150   &    -0.534   &      uy &   (6)\\
5398.279   &     4.4450   &    -0.630   &      uy &   (7) \\
5464.280   &     4.1430   &    -1.595   &      yy &   (6)\\
5466.396   &     4.3710   &    -0.630   &      uu &   (7) \\
5483.099   &     4.1540   &    -1.390   &      yu &   (3)\\
5522.446   &     4.2090   &    -1.419   &      uy &   (6)\\
5543.936   &     4.2170   &    -1.040   &      uy &   (7) \\
5546.506   &     4.3710   &    -1.124   &      yy &   (6)\\
5560.212   &     4.4340   &    -1.090   &      yy &   (7) \\
5618.632   &     4.2090   &    -1.250   &      yy &   (3)\\
5633.947   &     4.9910   &    -0.230   &      uu &   (7) \\
5636.696   &     3.6400   &    -2.511   &      uy &   (6)\\
5638.262   &     4.2200   &    -0.770   &      uy &   (7) \\
5651.469   &     4.4730   &    -1.763   &      uy &   (6)\\
5653.865   &     4.3870   &    -1.402   &      uy &   (6)\\
5661.346   &     4.2840   &    -1.756   &      yu &   (7) \\
5662.516   &     4.1780   &    -0.410   &      yy &   (3)\\
5679.023   &     4.6520   &    -0.756   &      yy &   (6)\\
5701.544   &     2.5590   &    -2.160   &      yy &   (7) \\
5705.465   &     4.3010   &    -1.355   &      yy &   (7) \\
5731.762   &     4.2560   &    -1.200   &      yy &   (7) \\
5741.848   &     4.2560   &    -1.672   &      yy &   (7) \\
5775.081   &     4.2200   &    -1.080   &      yy &   (3)\\
5778.453   &     2.5880   &    -3.430   &      yy &   (7) \\
5793.915   &     4.2200   &    -1.622   &      uy &   (6)\\
5811.914   &     4.1430   &    -2.333   &      uy &   (6)\\
5814.807   &     4.2830   &    -1.820   &      uy &   (6)\\
5855.077   &     4.6080   &    -1.478   &      yy &   (7) \\
5859.586   &     4.5490   &    -0.419   &      ny &   (7) \\
5861.110   &     4.2830   &    -2.450   &      ny &   (4) \\
5862.356   &     4.5490   &    -0.127   &      ny &   (7) \\
5902.473   &     4.5930   &    -1.797   &      uy &   (6)\\
5905.672   &     4.6520   &    -0.690   &      yy &   (7) \\
5916.247   &     2.4530   &    -2.914   &      yy &   (7) \\
5927.789   &     4.6520   &    -1.090   &      uy &   (4) \\
5929.676   &     4.5490   &    -1.211   &      uy &   (6)\\
5930.180   &     4.6520   &    -0.230   &      uy &   (7) \\
5934.655   &     3.9280   &    -1.170   &      uu &   (4) \\
5956.694   &     0.8590   &    -4.553   &      yy &   (7) \\
6027.051   &     4.0760   &    -1.089   &      yy &   (7) \\
6056.005   &     4.7330   &    -0.489   &      uy &   (6)\\
6065.482   &     2.6080   &    -1.470   &      yy &   (7) \\
6079.008   &     4.6520   &    -1.020   &      uy &   (7) \\
6094.373   &     4.6520   &    -1.566   &      uy &   (6)\\
6096.664   &     3.9840   &    -1.776   &      uy &   (6)\\
6151.618   &     2.1760   &    -3.299   &      yy &   (4) \\
6165.360   &     4.1430   &    -1.474   &      yy &   (4)\\
6173.336   &     2.2230   &    -2.880   &      yy &   (7) \\
6187.989   &     3.9430   &    -1.620   &      uy &   (7) \\
6200.313   &     2.6090   &    -2.405   &      yu &   (7)\\
6213.430   &     2.2230   &    -2.482   &      yy &   (4)\\
6226.736   &     3.8840   &    -2.120   &      uy &   (7)\\
6240.646   &     2.2230   &    -3.233   &      yy &   (4) \\
6270.225   &     2.8580   &    -2.573   &      yu &   (6)\\
6297.793   &     2.2230   &    -2.702   &      yy &   (7) \\
6315.811   &     4.0760   &    -1.610   &      yy &   (7) \\
6322.685   &     2.5880   &    -2.426   &      yy &   (4) \\
6481.870   &     2.2790   &    -2.984   &      yy &   (4) \\
6609.110   &     2.5590   &    -2.632   &      yu &   (6)\\
6627.544   &     4.5490   &    -1.475   &      uy &   (6)\\
6699.142   &     4.5930   &    -2.101   &      yy &   (7) \\
6705.101   &     4.6070   &    -1.057   &      ny &   (6)\\
6713.743   &     4.7960   &    -1.425   &      uy &   (6)\\
6739.522   &     1.5570   &    -4.794   &      yy &   (7)\\
6750.153   &     2.4240   &    -2.604   &      yy &   (7) \\
6810.262   &     4.6070   &    -0.986   &      yy &   (4)\\
6828.591   &     4.6380   &    -0.820   &      yy &   (7)\\
6842.685   &     4.6380   &    -1.169   &      uy &   (6)\\
6843.655   &     4.5490   &    -0.830   &      yy &   (7)\\
	   &		  &		&	  &				 \\
Fe II	   &		  &		&	  &				 \\
4993.358   &      2.8070  &    -3.684   &     yy  &   (8) \\
5197.568   &      3.2300  &    -2.293   &     yu  &   (6)\\
5234.625   &      3.2210  &    -2.280   &     yy  &   (9)\\
5264.812   &      3.2300  &    -3.130   &     yy  &   (9)\\
5325.553   &      3.2210  &    -3.320   &     yy  &   (10) \\
5337.722   &      3.2300  &    -3.338   &     yn  &   (6)\\
5414.073   &      3.2210  &    -3.580   &     yy  &   (9)\\
5425.257   &      3.1990  &    -3.390   &     yy  &   (10)\\
5991.376   &      3.1530  &    -3.650   &     uy  &   (10)\\
6084.111   &      3.1990  &    -3.881   &     uy  &   (10)\\
6149.246   &      3.8890  &    -2.719   &     uu  &   (6)\\
6247.557   &      3.8920  &    -2.430   &     uu  &   (10)\\
6369.462   &      2.8910  &    -4.230   &     yu  &   (10)\\
6432.680   &      2.8910  &    -3.570   &     yy  &   (10)\\
6456.383   &      3.9030  &    -2.190   &     uy  &   (10)\\
6516.077   &      2.8910  &    -3.279   &     yu  &   (6)\\
\hline
\end{longtable}
\tablefoot{ (1) \citet{2009A&A...502..989A}; (2) \citet{1981JPhB...14.4015S}; (3) \citet{2014ApJS..215...23D}; (4) VALD; (5) XXXO'Brian 1991; (6) \citet{2013A&A...555A.150T}; (7) {\sl Gaia}-ESO Survey linelist; (8) \citet{1998A&A...340..300R}; (9) \citet{2009A&A...497..611M}; (10) \citet{2003A&A...410..527B}.}
\end{longtab}

\twocolumn
\section{Fitting of the wings of the H$\alpha$ line to derive $T_{\rm eff}$}
\label{app:Halpha}

\begin{figure*}[t]
\begin{center}
\includegraphics[scale=0.35]{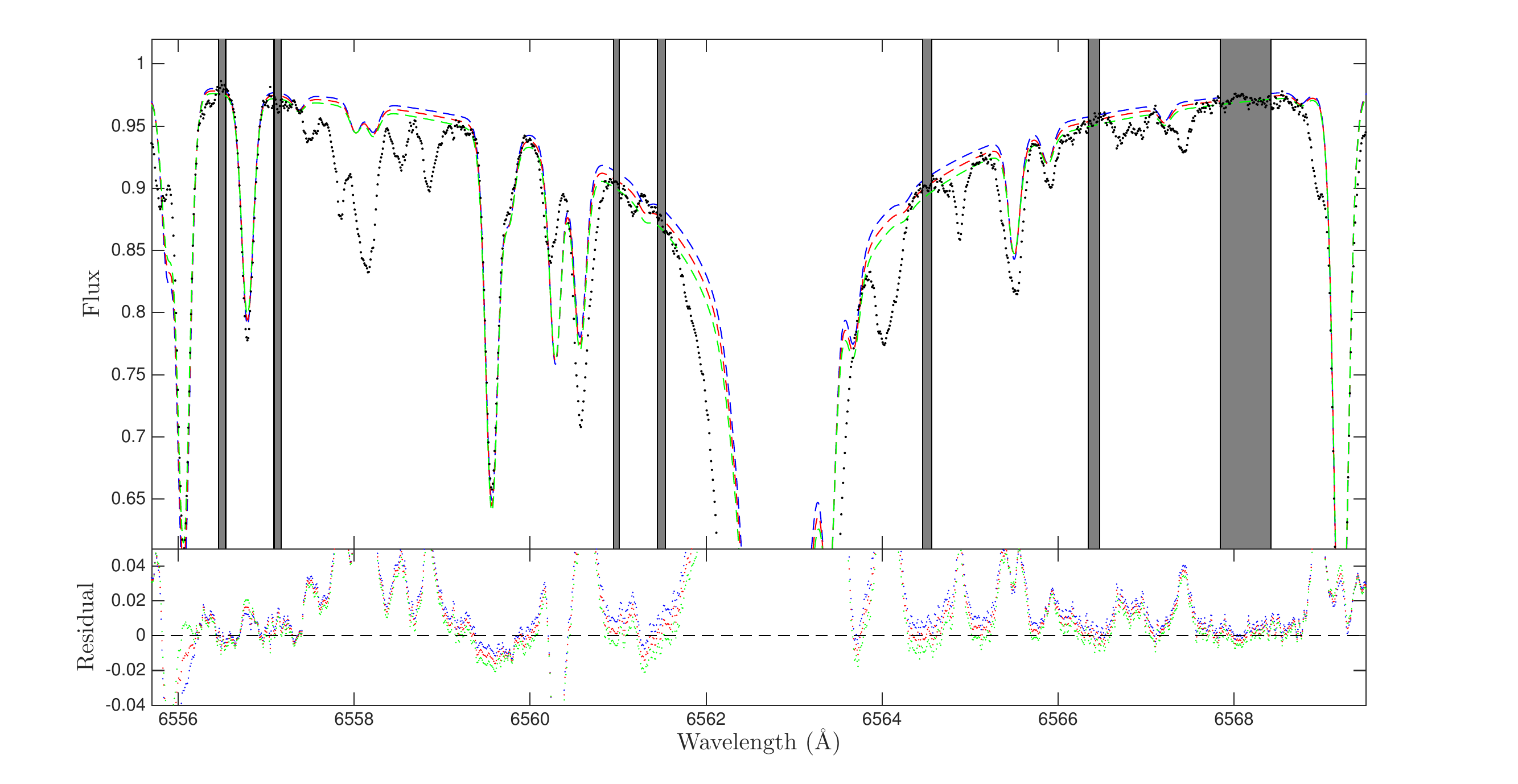}
\includegraphics[scale=0.35]{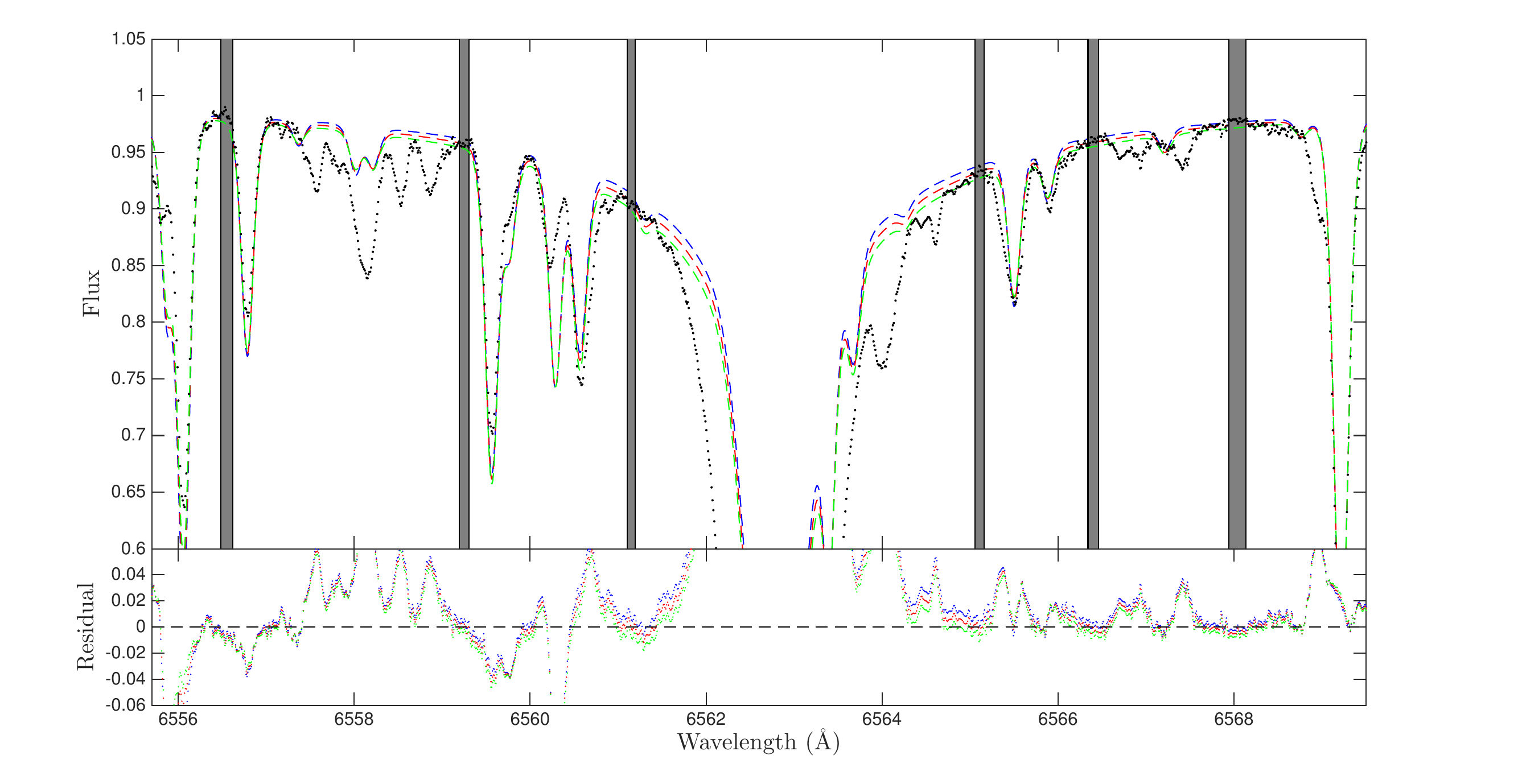}
\includegraphics[scale=0.35]{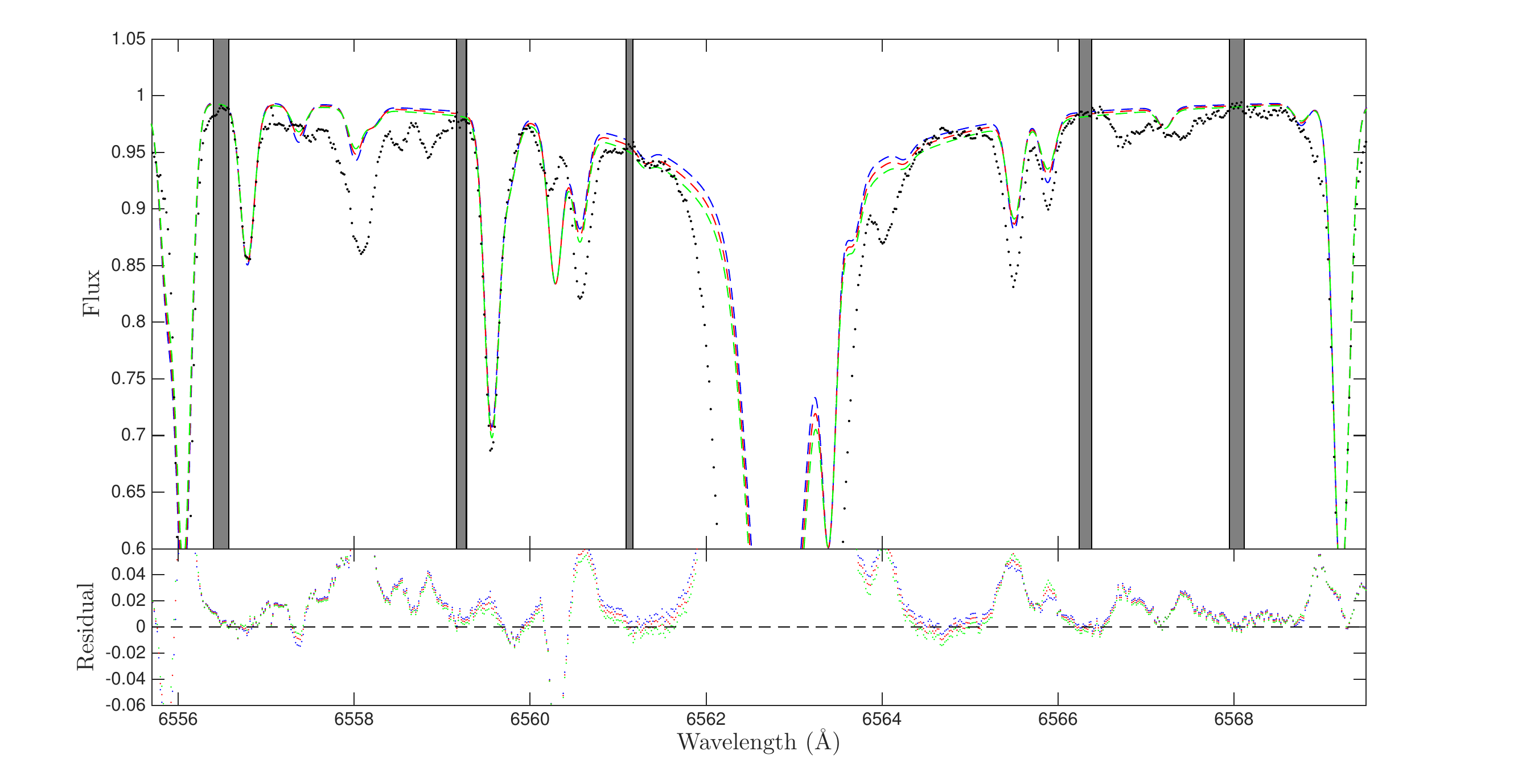}
\caption{Comparison of synthetic spectra with the wings of H$\alpha$
  for the {\sl Gaia} benchmark star $\epsilon$Vir ($T_{\rm eff} = 5056$\,K
  and $\log g=2.8$\,dex), $\xi$Hya ($T_{\rm eff} = 4991$\,K and $log
  g=$2.96), and HD\,107328 ($T_{\rm eff} = 4483$\,K and $\log
  g=2.0$\,dex). The middle synthetic spectrum (in red colour) is the
  best fit to the H$\alpha$ wings, and the other two synthetic spectra
  indicate the shape of the wings when $T_{\rm{eff}}$ is changed
  according to our estimated uncertainty. The marked grey regions were
  used to evaluate the goodness of the fit. Note that these are
  relatively short since many of the spectra have very little clean
  "line continuum" thanks to the cool temperatures which results in
  many molecular lines being present in the spectra.}
\label{app:halpha_BS}
\end{center}
\end{figure*}

\begin{figure*}[t]
\begin{center}
\includegraphics[scale=0.35]{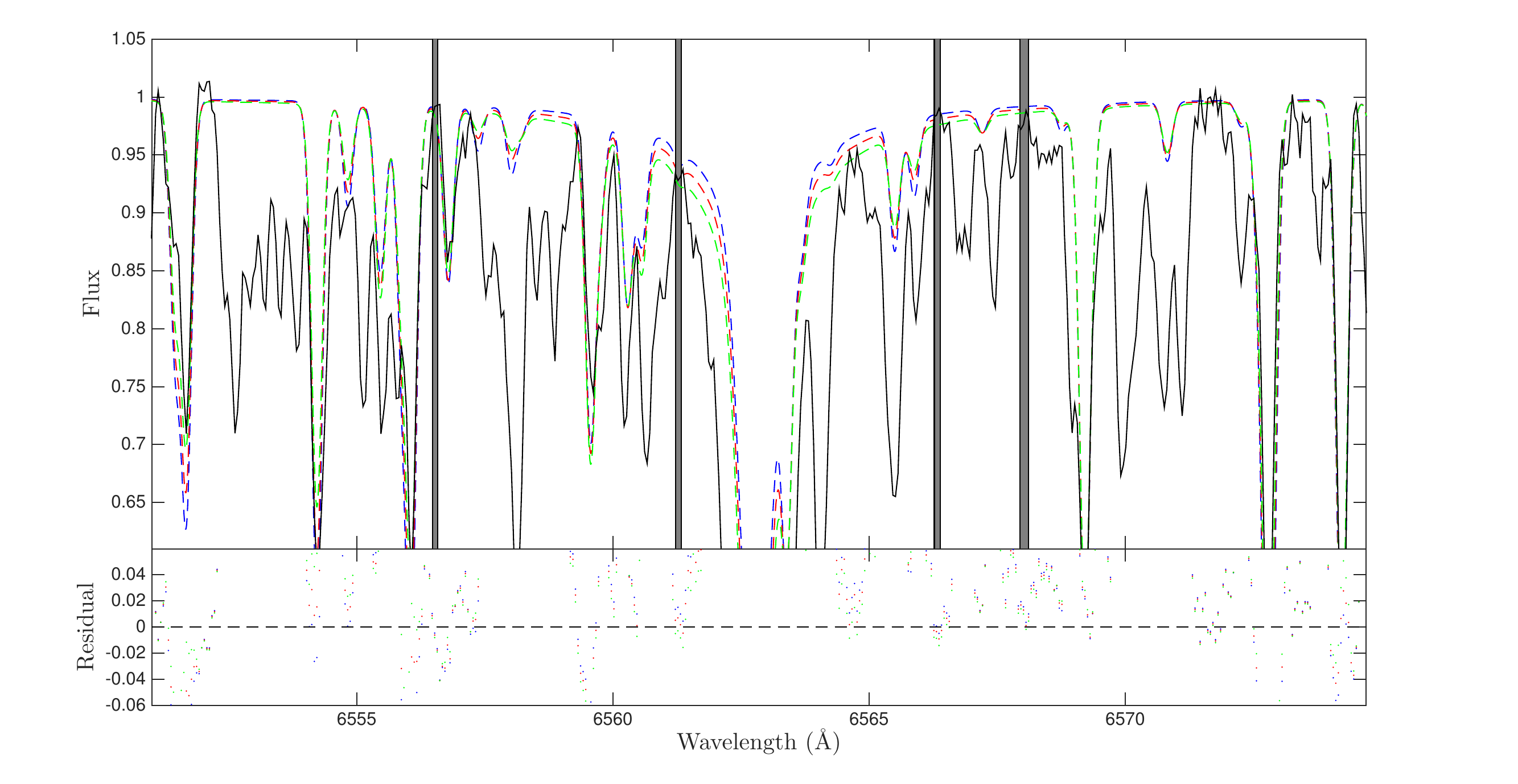}
\includegraphics[scale=0.35]{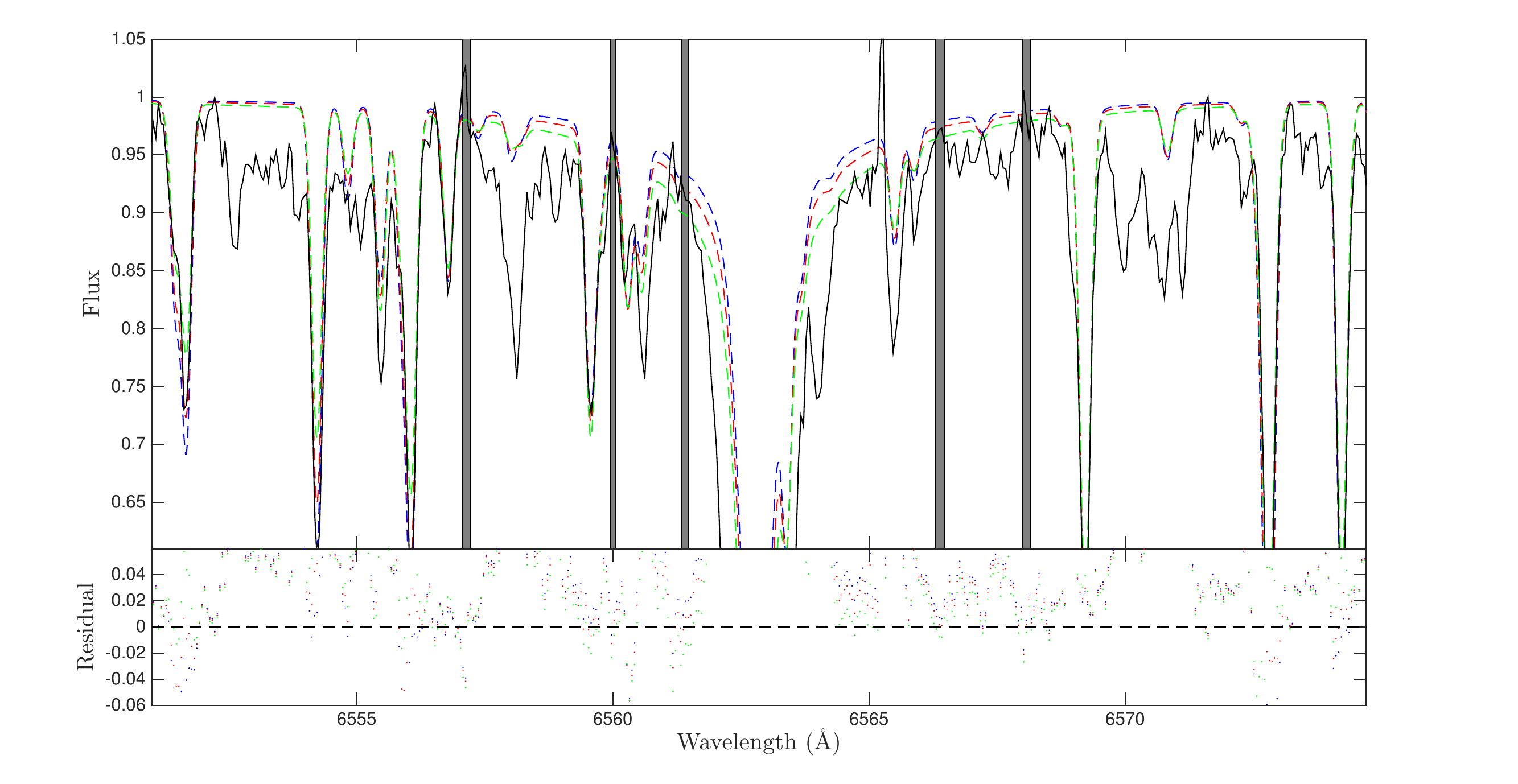}
\caption{Comparison of synthetic spectra with the wings of H$\alpha$
  two stars in NGC\,6528. Star-06 ($T_{\rm eff} = 4623$\,K and $\log
  g=1.98$\,dex) and Star-07 ($T_{\rm eff} = 4776$\,K and $\log
  g=2.50$\,dex). The middle synthetic spectrum (in red colour) is the
  best fit to the H$\alpha$ wings, and the other two synthetic spectra
  indicate the shape of the wings when $T_{\rm{eff}}$ is changed
  according to our estimated uncertainty. The marked grey regions were
  used to evaluate the goodness of the fit. Note the richer and more
  noisy spectra for these stars as compared to the benchmark stars
  displayed in Fig.\,\ref{app:halpha_BS}.}
\label{app_Halpa_6528}
\end{center}
\end{figure*}

In this study we have fitted the wings of the H$\alpha$ line by eye.
We chose this approach in order to gain a deeper understanding of the
issues facing the analysis of this line in the spectra of metal-rich
giant stars. However, for larger samples and for a more routine
approach an automated method should be attempted.

Figure\,\ref{app:halpha_BS} shows our fits for three benchmark stars
with different $T_{\rm eff}$ (see the Figure caption). The warmest
star is shown on top. This figure nicely illustrates the fact that the
wings of the H$\alpha$ line get less and less well developed in stars with  cooler and cooler temperatures. Eventually, the wings
become sufficiently weak that it is impossible to determine
$T_{\rm eff}$ from them.

Automated
routines for fitting the wings should ideally be used. This is not yet
well explored for metal-rich stars such as those in NGC\,6528. Methods
such as those in \cite{2002A&A...385..951B} should be possible to adapted but it
might become increasingly difficult to define suitable windows to fit
the spectra in metal-rich red giant stars. Figure\,\ref{app:halpha_BS}
shows the fits for three of the benchmark stars. As we are doing the 
fits by hand we have been able to customise the fitting regions. An experienced researcher may decide on different windows for the 
different spectra (as in our case). However, it would be feasible to  stick to a single set of windows with these high quality
spectra. This is more difficult when the
spectra have low S/N. Figure\,\ref{app_Halpa_6528}
shows how difficult it is to find clean regions in spectra with relatively
low S/N. The spectrum shown in the top panel has a S/N of 29 and the one
in the bottom panel has a S/N of 46. The two stars both have $T_{\rm eff}  > 4400$\,K, i.e. 
we would expect to be able to fit the H$\alpha$ line for both stars. 

\section{Surface gravity from the three Ca\,{\sc i} lines at 612.22, 616.22, and 643.91\,nm }
\label{app:logg}

\begin{table*}
\centering
\setlength{\tabcolsep}{1pt}
\caption{Surface gravities measured from the strong \ion{Ca}{i}  lines.}
\label{sgca}
\begin{tabular}{l|c|ccccccc}
      \hline
      \hline
  &    \multicolumn{7}{c}{Results from Method 2} \\
      Name  &$\log g_{\rm{r}}$ &$\log g_{\rm{Method2}}$ &$\log g_{1}$ &$\log g_{2}$ &$\log g_{3}$ &$\log g_{\rm{ave}}$ &$\sigma_{\log g_{\rm{ave}}}$ & Error\\
       \hline
      $\mu$Leo &2.51 &2.51 &2.61 &2.61 &2.41 &2.54 &0.09 &0.05\\
      HD107328 &2.09 &1.97 &2.26 &1.95 &2.20 &2.14 &0.13 &0.08\\
                           &---     &2.00 &2.18 &1.98 &2.19 &2.11 &0.10 &0.06\\
      $\beta$Gem &2.90 &2.89 &2.98 &2.91 &2.75 &2.88 &0.10 &0.06\\
      $\epsilon$Vir &2.77 &2.91 &2.93 &2.92 &2.64 &2.83 &0.13 &0.08\\
                               &---     &2.85 &2.93 &2.87 &2.70 &2.83 &0.10 &0.06\\
      $\xi$Hya &2.87 &2.94 &3.03 &2.96 &2.77 &2.92 &0.11 &0.06\\ 
\hline
\\
  &    \multicolumn{7}{c}{Results from Modified Method 2} \\
  \hline
      Arcturus & 1.64 &1.66 &1.77 &1.59 &1.68 &1.68 &0.07 &0.04\\
                           &--- &1.61 &1.77 &1.56 &1.55 &1.63 &0.10 &0.06\\
     $\alpha$Cet &0.68 &0.87 &0.61   &0.71 &1.19  &0.84 &0.25 &0.14\\
      $\gamma$Sge &1.05 &1.12 &1.18 &1.09 &0.92 &1.06 &0.11 &0.06\\
      $\alpha$Tau &1.11 &1.09 &1.13 &0.99 &0.80 &0.97 &0.13 &0.08\\
                              &---     &1.29 &1.47 &1.29 &0.87 &1.21 &0.25 &0.14\\
      $\beta$Ara &1.05 &2.05 &1.02 &2.05 &1.02 &1.36 &0.48 &0.28\\
                           &---     &2.06 &1.26 &2.06 &1.26 &1.53 &0.38 &0.22\\
      \hline
\end{tabular}
\tablefoot{The first column gives the name of the star. The second column gives the recommended $\log g$. The third column gives the obtained gravity by fitting the three Ca lines simultaneously in Method 2, and Modified Method 2. Log{\it g} determined  from each of the \ion{Ca}{i} lines are listed in columns 4 to 6. The average and standard deviation of the three values are given in column 7 and 8 and 
the error in the mean in column 9 ($\sigma/\sqrt{N_{\rm lines}}$).}
\end{table*}

In Method 2 and Modified Method 2 (Sect.\,\ref{sect:method2}) we fit
the three strong \ion{Ca}{i} lines at 612.22, 616.22, and 643.91\,nm simultaneously to determine $\log g$. This is not
necessarily the only way to make use of the gravity sensitivity of
these three lines. For example, it would be possible to fit each of
the lines individually and take the average of the three
determinations. Such an approach would also offer the possibility to
derive the scatter around the mean value as well as the error in the
mean. In Table~\ref{sgca} we report such results for the ten benchmark
stars we have studied.

We found that
for most stars, the typical uncertainty is around 0.10 dex. But, the
determined $\log g$ suffers from a larger uncertainty for a lower
gravity star. Except for $\beta$Ara, the average of the three measured
gravities ($\log g_{\rm{ave}}$) agree well with the obtained gravity
($\log g_{\rm{method2}}$) by fitting the three Ca lines simultaneously
listed in Table~\ref{sgca}. It is clearly shown that the determined
$\log g_{\rm{method2}}$ is strongly weighted by the second Ca line as
$\log g$ from the \ion{Ca}{i} 616.2\,nm line is much larger than the
other two lines.

\begin{figure*}
\begin{center}
\includegraphics[scale=0.4]{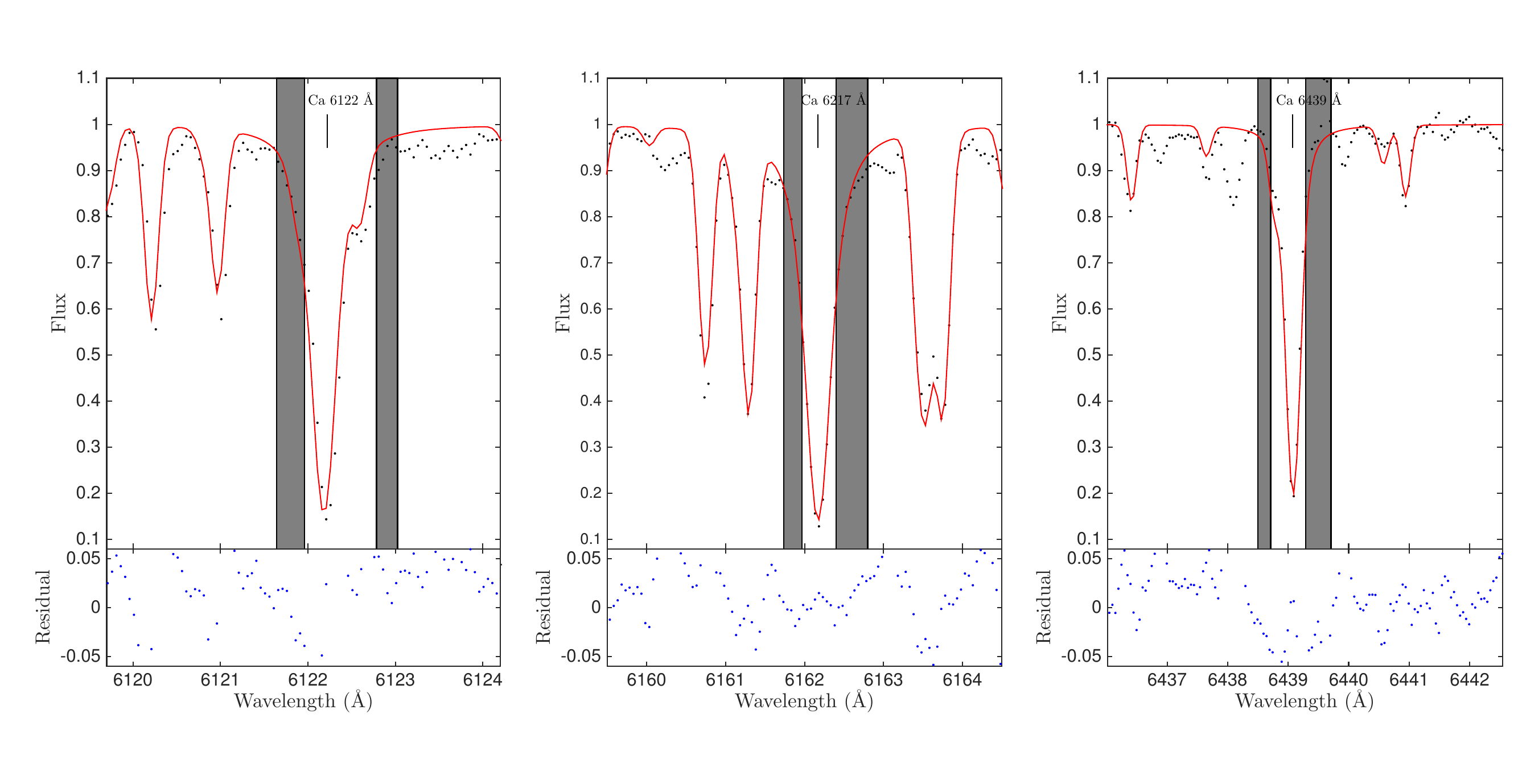}
\includegraphics[scale=0.4]{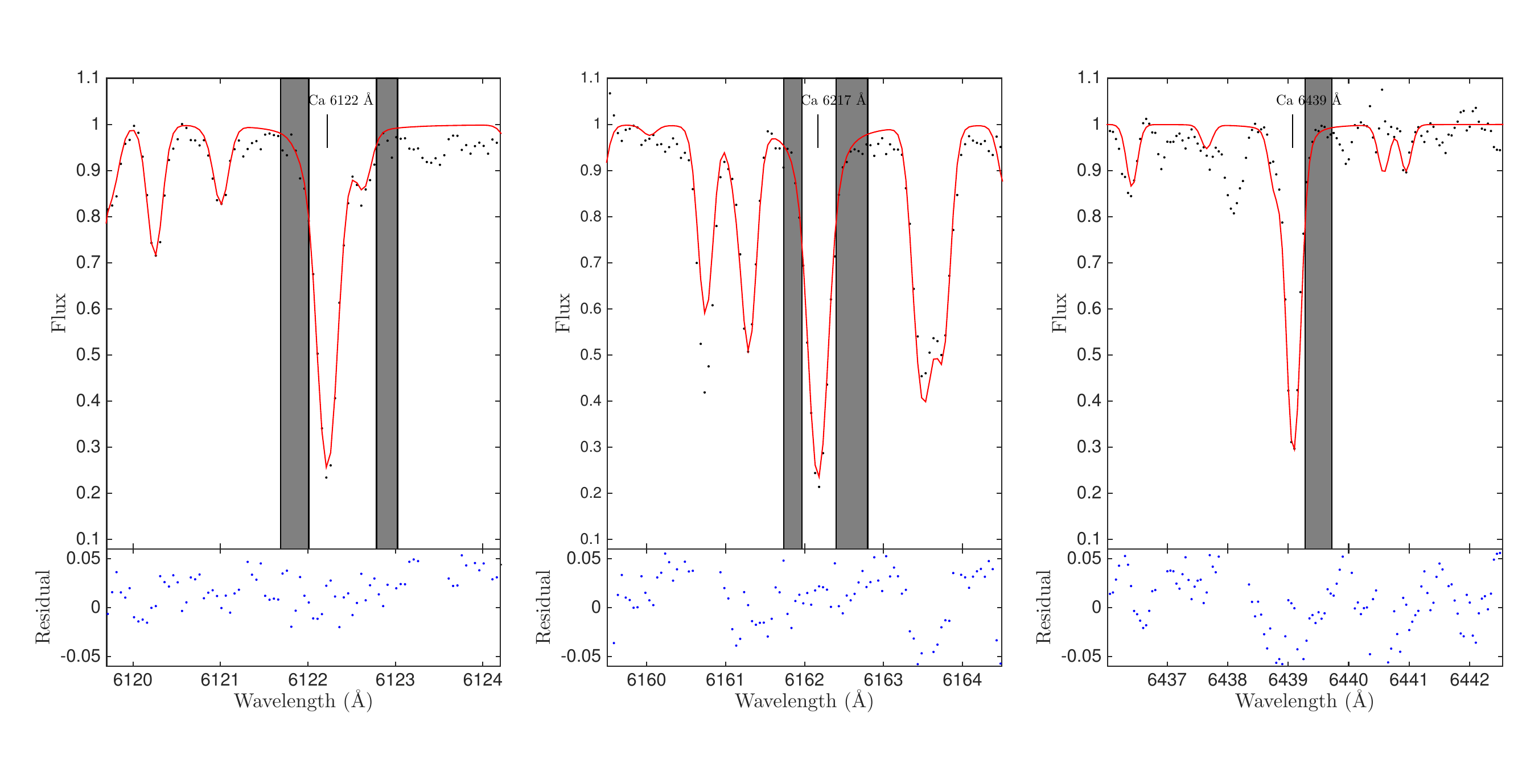}
\caption{Comparison of synthetic spectra fitted to the wings of the
  three strong \ion{Ca}{i} lines for two stars in NGC\,6528. Star-05
  ($T_{\rm eff} = 4277$\,K and $\log g=1.95$\,dex) and Star-07
  ($T_{\rm eff} = 4776$\,K and $\log g=2.50$\,dex). The observed
  spectrum is shown as a dotted line. The fitted spectrum is shown in
  red. This is the best fit as evaluated inside SME using the exact
  regions indicated with grey shaded regions.  All three lines are
  fitted simultaneously, as implemented in Method 2. In the lower
  panels the residuals are shown.}
\label{app_Cafit_6528}
\end{center}
\end{figure*}

\section{Exploration of impact of S/N on determination on stellar parameters for cool, metal-rich giant stars}
\label{app:SNR}

As discussed in Sects.\,\ref{sect:snrsb} and \ref{sect:SNR6528} the
S/N of the spectra influences the results, at least for low S/N. This is one
of the major findings of this study. 

\begin{figure*}
\begin{center}
\includegraphics[trim = 30mm 10mm 30mm 5mm, clip, width=18cm]{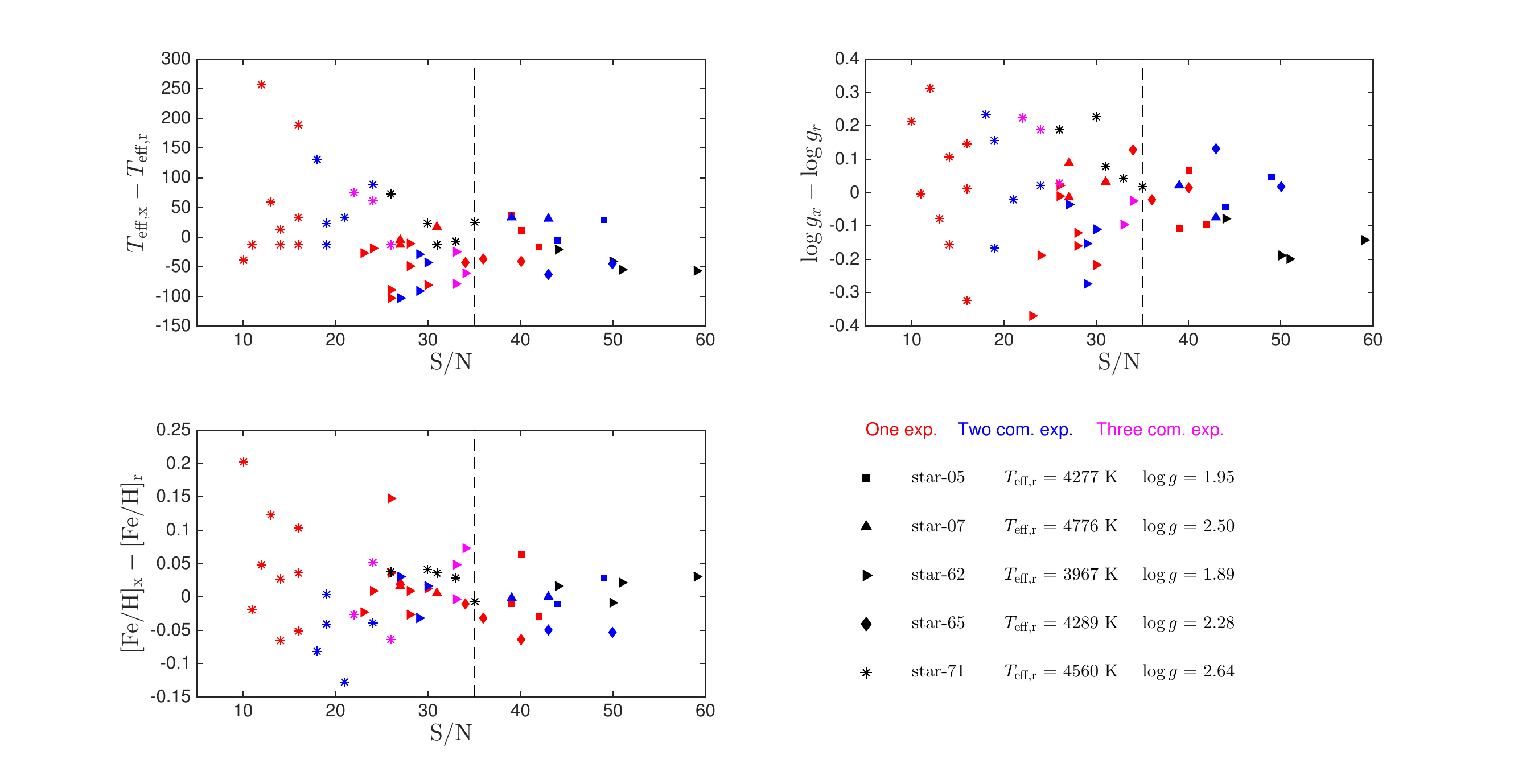}
\caption{Comparison of the derived stellar parameters as a function of
  S/N. The $y$-axes show the difference between the value for each
  parameter derived from the low S/N spectrum and the value derived
  from the final combined spectrum (i.e., that used in the analysis in
  Sect.\,\ref{sect:metallicity}. The stellar parameters of the five
  stars are given in the legend. The colours refer to how many spectra
  have been combined to obtain the analysed spectrum (colours as
  indicated in the legend).}
\label{figapp:SNR}
\end{center}
\end{figure*}

For several of the stars in NGC\,6528 we have multiple exposures. This
allows us to analyse spectra of different S/N for the same star.
Using Method 2 and Modified Method 2 we analysed the spectra of five
stars (Star-05, -07, -62, -65 and -67, IDs as in Table\,\ref{gcs}).
Figures\,\ref{msnr} and \ref{figapp:SNR} show the results of our
investigation. As a  reference value we use the final parameters
determined for each star, Table\,\ref{sp}.
The results shown for [Fe/H] were used to create Fig.\,\ref{msnr}.

It is easily seen, as expected, that when we analyse spectra of lower
and lower S/N the scatter increases for all three parameters. It is in
particular acute for $\log g$ which blows up to a rather large error
quickly as the S/N deteriorates.

For $T_{\rm eff}$ there is a marked increase in scatter and different
stars appear, perhaps to behave in different ways.  It is in
particular interesting to compare and contrast star-62 and star-71.
For star-71 (with $T_{\rm eff}$ of 4560\,K, thus analysed with Method
2) we systematically underproduce $T_{\rm eff}$ when we go to lower
S/N, while the opposite is true for star-62 (with $T_{\rm eff}$ of
3967\,K, thus analysed with Modified Method 2). For $T_{\rm eff}$
there is thus a difference between Method 2 and Modified Method 2. We
recall that Method 2 uses the wings of the H$\alpha$ line to obtain
$T_{\rm eff}$ whilst Modified Method 2 uses excitation equilibrium of
\ion{Fe}{i} lines as the H$\alpha$ wings
disappear. Figure\,\ref{figapp:SNR} shows the detailed results.

The results for [Fe/H] are discussed in detail in the main body of the paper.

\end{appendix}

\end{document}